\documentclass[sigconf]{acmart}

\AtBeginDocument{%
  \providecommand\BibTeX{{%
    \normalfont B\kern-0.5em{\scshape i\kern-0.25em b}\kern-0.8em\TeX}}}

\copyrightyear{2024}
\acmYear{2024}
\setcopyright{acmlicensed}\acmConference[CHI '24]{Proceedings of the CHI
Conference on Human Factors in Computing Systems}{May 11--16,
2024}{Honolulu, HI, USA}
\acmBooktitle{Proceedings of the CHI Conference on Human Factors in
Computing Systems (CHI '24), May 11--16, 2024, Honolulu, HI, USA}
\acmDOI{10.1145/3613904.3642827}
\acmISBN{979-8-4007-0330-0/24/05}

\acmSubmissionID{930}

\begin{document}

\title[Charting the COVID Long Haul Experience]{Charting the COVID Long Haul Experience - A Longitudinal Exploration of Symptoms, Activity, and Clinical Adherence}

\author{Jessica Pater}
\orcid{0000-0001-5115-8439}
\email{jessica.pater@parkview.com}
\affiliation{
  \institution{Parkview Research Center}
  \city{Fort Wayne}
  \state{IN}
  \country{USA}}

\author{Shaan Chopra}
\orcid{0009-0002-6471-1031}
\email{schopra7@cs.washington.edu}
\affiliation{
  \institution{University of Washington}
  \city{Seattle}
  \state{WA}
  \country{USA}}

\author{Juliette Zaccour}
\orcid{0009-0009-0973-7168}
\email{juliette.zaccour@mail.utoronto.ca}
\affiliation{
  \institution{University of Toronto}
  \city{Toronto}
  \state{Ontario}
  \country{Canada}}

\author{Jeanne Carroll}
\orcid{0000-0002-4977-2250}
\email{jeanne.Carroll@parkview.com}
\affiliation{
  \institution{Parkview Research Center}
  \city{Fort Wayne}
  \state{IN}
  \country{USA}}

\author{Fayika Farhat Nova}
\orcid{0000-0002-2606-1958}
\email{fayikafarhat.nova@parkview.com}
\affiliation{
  \institution{Parkview Research Center}
  \city{Fort Wayne}
  \state{IN}
  \country{USA}}

\author{Tammy Toscos}
\orcid{0000-0002-9679-5359}
\email{tammy.toscos@parkview.com}
\affiliation{
  \institution{Parkview Research Center}
  \city{Fort Wayne}
  \state{IN}
  \country{USA}}

\author{Shion Guha}
\orcid{0000-0003-0073-2378}
\email{shion.guha@parkview.com}
\affiliation{
  \department{Faculty of Information and Department of Computer Science}
  \institution{University of Toronto}
  \city{Toronto}
  \state{Ontario}
  \country{Canada}}
\affiliation{
  \institution{Parkview Research Center}
  \city{Fort Wayne}
  \state{IN}
  \country{USA}}

\author{Fen Lei Chang}
\orcid{0000-0003-4415-7483}
\email{fenlei.chang@parkview.com}
\affiliation{
  \institution{Parkview Health}
  \city{Fort Wayne}
  \state{IN}
  \country{USA}}

\renewcommand{\shortauthors}{Pater et. al.,}

\begin{abstract} 
COVID Long Haul (CLH) is an emerging chronic illness with varied patient experiences. Our understanding of CLH is often limited to data from electronic health records (EHRs), such as diagnoses or problem lists, which do not capture the volatility and severity of symptoms or their impact. To better understand the unique presentation of CLH, we conducted a 3-month long cohort study with 14 CLH patients, collecting objective (EHR, daily Fitbit logs) and subjective (weekly surveys, interviews) data. Our findings reveal a complex presentation of symptoms, associated uncertainty, and the ensuing impact CLH has on patients' personal and professional lives. We identify patient needs, practices, and challenges around adhering to clinical recommendations, engaging with health data, and establishing "new normals" post COVID. We reflect on the potential found at the intersection of these various data streams and the persuasive heuristics possible when designing for this new population and their specific needs. 
\end{abstract}



\received{14 September 2023}
\received[revised]{12 December 2023}
\received[accepted]{19 January 2024}

\maketitle

\section{Introduction}
COVID-19 is no longer a pandemic -- it has become endemic in the United States and across the world \cite{CDC_end}. For many people, this illness will resolve and not have any lasting effects. Unfortunately, for an estimated one-third of those infected, long-term, chronic issues will persist. This condition is known as Post-Acute Sequelae of COVID-19, or more conventionally, COVID Long Haul (CLH). Common CLH symptoms include brain fog, fatigue, chest pain, cough, and headaches \cite{CDC_CLH}. What makes this chronic illness even more challenging is that it is not necessarily a continuation of one's current symptoms, but could also entail the onset of new symptoms. 

The emerging nature of this health condition means that researchers are rapidly mobilizing, examining everything from clinical pathology \cite{Mehandru} to new treatments \cite{Yong} and susceptibility to CLH \cite{Marx}. Researchers are also focusing on understanding specific aspects, such as hospitalization for COVID \cite{Karaarslan} or conditions such as Chronic Obstructive Pulmonary Disease (COPD) \cite{Vanichkachorn} or diabetes \cite{Mehandru}, for their potential impact as risk factors for developing CLH. Computational research has expanded to assess the viability of detection \cite{Zhu22}, enhanced modeling \cite{Antony} and better understand how social media supports information-seeking related to CLH \cite{Pater23}. However, lived experience of individuals with CLH, including the variation in symptoms and severity of symptoms in daily life, remain less understood. 

Naturalistic settings differ from other study approaches (e.g., lab or clinical settings) in the ability to explore different contexts such as the workplace, home, or social settings. The HCI community is uniquely positioned to not only research lived experiences through technological approaches but also to translate these findings into design considerations for future technologies to better support individuals with CLH.  

By designing studies that take into consideration multiple data streams in addition to individuals' perceptions, needs, and lived experiences, we can paint a more holistic picture for the sociotechnical ecosystem in which a CLH patient lives. Assessing  sociotechnical systems that people live in has a long history in CSCW and HCI research, in technology design~\cite{Miller,Barricelli}, decision-making~\cite{Bossen} as well as evaluation of research environments~\cite{Murnane,Siek}. Moreover, framing the research through a lived experience lens affords a deep contextualization of objective measures. Objective data, such as validated survey and clinical measures, can thus be contextualized with subjective data collected through participant engagements (e.g., interviews, focus groups, observations) to generate a more complete understanding of environmental or behavioral impacts on an individual's CLH symptoms. This is not possible through the analysis of electronic health records (EHRs) on its own \cite{Wang,Pater2019}. 

To meet this challenge, we designed a longitudinal study that followed 14 participants for 3 months - February through April of 2022. Participants were patients in a specialty Post-COVID Clinic. Our findings are triangulated from four data streams which represent objective (EHR, Fitbit Logs) and subjective (weekly surveys, exit interview) data. To provide a starting point for design recommendations for technology to support this chronic illness, we took the approach of collecting data from various parts of a participant's life to explore what could be measured and how patients are currently experiencing and engaging with data. The data collected helps us to better understand the evolving nature of patients' CLH symptoms over time (surveys), track key indicators related to their most common CLH symptoms (activity and sleep), and correlate what actions or behaviors patients self-reported with what was verifiable within their EHR. Due to the embedded nature of the research, certain key clinical insights and validation were possible.

Through this research, we make three contributions. First, we provide a contextually rich description of the CLH patient experience and sociotechnical environment. Participants described the needs and challenges of establishing a ``new normal'' as their CLH symptoms evolved, mixed findings with respect to engaging with their health tracking data, and various challenges with respect to adherence to clinical recommendations. We also found conflicting information between what participants reported and what was found within the EHR. Second, triangulating between different data streams, we discuss pros and cons of each and the potential of combining different data streams to better characterize patient experiences with this emerging chronic illness. Lastly, we engage with HCI and health informatics literature to provide design heuristics for supporting and understanding patient experiences with CLH.

\section{Related Work} 
\subsection{Defining the Domain: COVID Long-Haul (CLH)}
CLH is an emerging chronic illness that lacks clear diagnostic criteria and thus definition \cite{Phillips}. A recent World Health Organization Delphi study, with inputs from almost 200 clinicians from across the world, crowd sourced a functional definition for CLH: \textit{``condition [which] occurs in individuals with a history of probable or confirmed SARS-CoV-2 infection, usually 3 months from the onset, with symptoms that last for at least 2 months and cannot be explained by an alternative diagnosis''}~\cite{Soriano}. The presentation of CLH varies by the systems it impacts (e.g., respiratory, neurological, cardiovascular, etc) as well as the different types of symptoms that are manifested (e.g. brain fog, dyspnea (commonly known as shortness of breath), fatigue, etc.) with one study documenting 73 unique symptoms \cite{Mehandru}. The emerging data shows that the global estimates of CLH prevalence is 0.43, where the U.S. showed a prevalence of 0.31 (95\% CI, 0.14-0.57) \cite{Chen}. CLH is present in patients who experienced varying levels of COVID disease, from asymptomatic to those who were hospitalised and on a ventilator \cite{Townsend}. Recently the NIH \cite{Collins} and CDC \cite{Raths} have launched programs to fund extensive research about this evolving illness. Although the HCI community has seen research into a post-COVID world \cite{Troiano,Shi,Rao,Last}, only some deal directly with CLH symptoms or patients' lived experiences \cite{Corman,Homewood}.

\subsection{Chronic Illness, HCI, \& Wearable Technology}
The HCI community has consistently studied chronic illnesses over the last several decades. Early research examined the role of computational approaches in coping strategies of patients suffering from various chronic illnesses \cite{Cubranic,Bers}, the use of technology for information and communication \cite{Mankoff,Sun,Pang,O'Kane}, and home-based supports \cite{Birnholtz,Taylor}, to name a few. More recent scholarship has focused on self-management \cite{Cha,Mead}, finding information online \cite{Karusala}, interactions with personal health records \cite{Salamah}, and engaging with health data tracking devices \cite{Ding}. As CLH continues to evolve, the HCI community is in a unique position to utilize tools and approaches to better explore this emerging chronic illness.

Several studies have examined information sharing surrounding chronic illness. For example, Pang et. al., examined technologies that people preferred for sharing information and their information sharing routines, with privacy being a central concern \cite{Pang}. Key findings included that sharing information about chronic illness was preferred on mobile platforms and that communication around chronic illness differed from that of typical family interactions \cite{Pang}. MacLeod et. al., explored design considerations surrounding people with rare diseases and found that patients often found communicating with both people in their personal lives and their doctors difficult \cite{MacLeod}. Additionally, their findings about technology having a function in various roles--including advocacy, copers, researchers, and record keepers--are especially important with respect to CLH patients as, even though it is still emerging, there are  examples of technology design and development serving in these roles for CLH patients \cite{Dalko,Khondakar}. Hong et. al. looked at the use of diaries to better understand patients' chronic illness experiences and found that technology could be beneficial to help elucidate both physical and emotional experiences within the context of their everyday lives \cite{Hong}. Although these studies highlight various aspects needed for better understanding the holistic impact of a chronic illness, they are somewhat limited by the data they were able to collect for analysis.

Activity trackers have been used in research focused on chronic illness because they are a convenient commercial off-the-shelf way to assess issues such as activity and sleep levels for those dealing with cancer \cite{Rossi}, diabetes \cite{Heuschkel}, and chronic fatigue syndrome \cite{Roche}. With research on the self-reports of sleep being mixed -- showing that self-reports are both a legitimate vehicle for measuring accurate sleep \cite{Ibáñez} but also can sometimes be biased \cite{Taylor84} -- technology can provide a cost-effective way to objectively measure sleep and activity levels. Despite questionable accuracy of sleep data recorded by Fitbits, they provide appropriate levels of insight for studies where they are not being used for diagnostic purposes \cite{Haghayegh}. When 
enhanced specificity is not important, the Fitbit and other similar accessible sleep/fitness tracking devices can provide meaningful, objective data that is useful both clinically and as a feedback tool for patients.

\subsection{Clinical Data \& EHR}
EHRs are the digitization of a patient's clinical chart, a repository of their health-related data \cite{Clynch15}), and have revolutionized healthcare by enabling true collaborative work across various healthcare settings \cite{Bowman13}. Recently, data from consumer health technologies like activity trackers and implanted devices have also been directly integrated into the EHR \cite{Chokshi}. Several studies have found that integration of activity trackers into the EHR gives providers and patients a way to not only track key health issues like sleep, activity, and heart rate but also can serve as a motivation for patients to engage with their health data \cite{Lv,Plastiras}.

The broader HCI community has a vast array of patient/provider-centered research related to the EHR from workflows \cite{Marthe21,Tang15} to data sharing \cite{Oh22,Murphy17} and collaboration \cite{Veinot10,Bossen12}. Pater et al. provide an example of how integrating EHR data with other patient reported outcomes provide unique insights and allow for a more complete analysis of clinical and non-clinical health indicators \cite{Pater2019}. This approach of triangulating knowledge across various data streams was adapted for this research. The ability to access various streams allows for a more holistic understanding of the phenomenon and its impact on patients' lives, connecting critically rich subjective contexts to objective clinical data \cite{Cadmus,Kawu}.

The HCI community has a deep history of research integrating EHRs and EHR data. From the design standpoint, research has focused on designing dashboards both within \cite{Iott} and outside of \cite{Yoo,Verbert} the EHR and the usability of the EHR \cite{Marks16}. Other studies have focused on assessing specific aspects of data acquisition and use, such as understanding clinical workflows \cite{Tang15,Marthe21} and dynamics between stakeholders\cite{Cajander19}. Researchers have leveraged this understanding in trying to connect external, subjective data with objective health data and outcomes documented within the EHR \cite{Pater19,Munmun} and speculating on how HCI research could potentially be integrated into future work \cite{Pater21,Haldar20}. The HCI and health community has also examined impacts of the EHR, such as enhanced collaborative work within clinical spaces \cite{Bossen12, Veinot10}. Additionally, research has focused on the EHR and how it supports enhanced data sharing \cite{Murphy17,Oh22}, a critical aspect for decision making \cite{Zhang21}.

Thus, when trying to define this new clinical issue, the EHR is a central component in collecting and sharing objective health measures. However, it only provides a limited snapshot of health at a given point, and lacks contextually-rich power to understand the lived impact of the illness on individuals. This led to the research design of integrating objective and subjective data streams to get a better, holistic understanding of this emerging chronic health issue in addition to affording us the ability to cross-validate data elements that were collected from both types of data streams.

\subsection{Data Triangulation}
Data triangulation is a technique where multiple methods or data are used to develop a comprehensive and/or contextually rich understanding \cite{Patton}. Triangulation is a common and productive research strategy in qualitative studies, including triangulating data from different types of data sources \cite{Carter}. The mixing of qualitative methods and their data sources for analysis allows for the uncovering of different perspectives \cite{Morse}, which is often critical in emerging phenomena \cite{Fusch}. For the purposes of this research study, we draw from Cohn and Manion's work that focuses on triangulation as a way to explain the complexity of human behavior\cite{Cohen}.

Within the HCI domain, data triangulation is a practice that has taken root within the community over the last decade. This technique has uncovered gaps between theory and practice, especially with respect to behavior change \cite{Chaudhry,Mamykina,Rogers}, phenomenological understanding \cite{Vedant}, and the analytical power it provides researchers \cite{Maestre,Pater2019}. Within the health space, Zhu et. al., found that data triangulation and communication could positively impact workflow inefficiencies and burdens during prehabilitation and pre-operative appointments \cite{Moffa}. Ernala et. al., used data triangulation within patient data of schizophrenic patients to assess the efficacy of proxy diagnostic signals in identifying a potential diagnosis. They found that using signals developed from social media, which had high internal validity, had very poor external validity when used on clinical data \cite{Ernala19}. These examples provide what can be accomplished with data triangulation, in this case interview data and a combination of self-reported and social media data respectively. Pater et. al., combined both qualitative and quantitative data approaches to challenge common assumptions related to eating disorder content on social media. Through the interview process, they uncovered the deep impacts of social media content on participants' disease, but just looking at their social media history or the surveys they took as part of the study, this would not be apparent \cite{Pater19}. Thus, using data validation within HCI studies provides a method where complex data collection can be parsed to obtain deeper, more holistic knowledge.

\section{Methods}
This research study is part of a larger initiative focused on researching the presentation and impact of CLH on patients' everyday health and wellness \cite{Pater23,Fisher,Bohn,Todd}. This qualitative study had several concurrent components. Participants were enrolled in the study for up to 24 weeks. This included a 12-week period of wearing a Fitbit and taking weekly surveys and then a culminating interview. Data streams included daily Fitbit data (including sleep and activity tracking data), weekly surveys, EHR data, and end-of-study interviews. Participants spent approximately two hours in the study (5 min per survey x 12 surveys plus a 60 minute interview). Participants were compensated by keeping their Fitbits after the end of the study in addition to a \$20 e-giftcard after their interview was completed. This study was reviewed and approved by the Parkview Health Institutional Review Board.

\subsection{Clinical Setting}
The Parkview Post-COVID Clinic (PPCC) is an integrative specialty clinic embedded in a health system in the Midwest, United States. The PPCC opened to patients in March 2021, making it one of the first 66 post-COVID clinics in the U.S. \cite{COVID}. At the time of publication, the clinic has a panel of over 900 patients. The PPCC team includes clinicians from the fields of neurology, physical medicine and rehabilitation, neuropsychology, pharmacy, and physical therapy. The clinical and research team have a standing weekly meeting to discuss research and data needs. A unique aspect to the PPCC is the patient-reported data that is collected prior to each patient's initial visit. Each patient is requested to provide their top 5 symptoms ranked by severity, a timeline of their symptoms, impact on their daily life, and what their best/worst CLH symptom day looks like. The research team is a collaboration between PPCC clinical leadership and a scientific team from an embedded research unit within the health system.

\subsection{Participants}
A total of 15 patient participants were recruited through the Parkview Post COVID Clinic (PPCC). A research nurse reviewed records of patients who had recently been seen for the first time in the PPCC. If the inclusion criteria were met, the nurse made recruitment calls to the patients, excluding patients who stated in their personal preferences within the EHR that they did not wish to be contacted. Potential participants were contacted in January 2022 and enrolled by February 2022. Inclusion criteria included: Age 18 or older, patient of the PPCC with an initial consultation within 30 days prior to their enrollment, able to speak/read/write in fluent English, able to understand all aspects of the study and provide informed consent, have an active email address, and have access to a smartphone/computer/tablet. 

Of the 15 participants recruited, 14 completed the study. P18 was removed for lack of participation as they stopped responding to surveys during week 5. P21 did not participate in the interview portion of this study as they stopped communicating with the research team, however since they completed 11/12 weekly surveys we chose to include their data for analysis. Table \ref{tab:demo} highlights the demographic data of the study participants.

\begin{table*}[ht]
\begin{tabular}{llllccc}
\textbf{\begin{tabular}[c]{@{}l@{}}Participant\\ ID\end{tabular}} & \textbf{Gender} & \textbf{\begin{tabular}[c]{@{}l@{}}Age\\ Range\end{tabular}} & \textbf{Race} & \multicolumn{1}{l}{\textbf{BMI}} & \multicolumn{1}{l}{\textbf{\begin{tabular}[c]{@{}l@{}}COVID \\ Hospitalization\end{tabular}}} & \multicolumn{1}{l}{\textbf{\begin{tabular}[c]{@{}l@{}}Comorbidity \\ Burden\end{tabular}}} \\ \hline
P10 & Male & 71-80 & White/Caucasian & 24.5 & Yes & 1 \\
P11 & Female & 41-50 & White/Caucasian & 27.0 & No & 1 \\
P12 & Male & 51-60 & White/Caucasian & 35.7 & No & 0 \\
P13 & Male & 51-60 & White/Caucasian & 31.3 & No & 1 \\
P14 & Female & 71-80 & White/Caucasian & 32.0 & No & 1 \\
P15 & Female & 41-50 & White/Caucasian & 42.3 & No & 0 \\
P16 & Male & 41-50 & Black/African American & 33.8 & No & 2 \\
P17 & Female & 51-60 & White/Caucasian & 37.0 & No & 0 \\
P19 & Female & 41-50 & White/Caucasian & 40.7 & Yes & 1 \\
P20 & Female & 31-40 & White/Caucasian & 34.2 & No & 1 \\
P21 & Female & 31-40 & White/Caucasian & 30.3 & No & 0 \\
P22 & Female & 51-60 & White/Caucasian & 47.7 & Yes & 0 \\
P23 & Male & 31-40 & Black/African American & 43.3 & No & 2 \\
P24 & Female & 41-50 & White/Caucasian & 31.4 & No & 1
\end{tabular}
\caption{Participant demographics and health information}
\label{tab:demo}
\end{table*}

\subsection{Data Collection}
\subsubsection{EHR Data} 
The research team had access to the EHRs of the participants. The EHR system used in the clinic, Epic, was used to assess if patients met the study's inclusion criteria and to collect enrolled participants' demographic data. Additionally, the study team reviewed each participant's initial PPCC progress note to obtain their top five CLH symptoms reported to the clinic and the recommendations/care plan provided to patients by the PPCC. Documentation of this data is standard-of-care for every patient seen within the PPCC. The study team also reviewed the health system encounters of the participants after their first PPCC visit to validate if participants were following clinical recommendations. Team members who did these chart reviews completed Epic's ambulatory nurse training and all data presented here has been de-identified in compliance with HIPAA regulations. Participants were aware of this access as it was part of their informed consent.

\subsubsection{Weekly Surveys} Participants were sent a survey each week via HIPAA-compliant Survey Monkey. The survey comprised of patient reflections on data taken from the EHR. The weekly surveys asked about the following: their top 5 symptoms and how each ranked in severity, if symptoms had gotten better/worse/stayed the same since the last survey, adherence/progress being made towards care plan recommendations, and assessment of overall health (see Appendix A). The top five symptoms were as patients initially reported to the clinic (taken from the EHR) which were used for the baseline in the survey or as they reported in their weekly surveys. Thus, similarities in symptoms like brain fog and concentration are present in the results. Care plans included both lifestyle-related recommendations and clinical recommendations tailored to each participant's main issues. All participants were given the same lifestyle recommendations which included diet/nutrition/hydration, exercise, sleep hygiene, stress management, weight management, and PPCC support group participation. All clinical recommendations were patient-specific (e.g., MRI of brain, physical therapy, etc.) with the exception of medication adherence and COVID vaccination. In total, there were over 50 unique recommendations within the care plans with each participant assigned 10 to 21 recommendations. Surveys were sent out every Monday morning and participants were contacted up to three times via email/text/phone call if their surveys were not completed by Wednesday morning of each week. 10 out of 14 participants completed all 12 weekly surveys. A total of 156 surveys were collected for a compliance rate of 92.9\%.

\subsubsection{Fitbit} Participants were mailed a Fitbit Luxe device and were asked to wear it daily for the duration of the study to track their step count and sleep data. Individual study emails were set up and attached to individual devices so that the study team would have access to the data logs. On completion of the study, participants were sent detailed instructions on how to transfer the Fitbit account from the study email to their personal email as the technology was part of their compensation.

\subsubsection{Interviews} The research team conducted a total of 13 semi-structured interviews with individual participants at the completion of the study period. The goal of the interview was to contextualize each participant's health tracking, survey, and EHR data to get a holistic understanding of their CLH experiences. There were several themes to the questions. Participants were asked about their weekly surveys with a focus on symptom reporting frequency, if they were able to share as much information about their symptoms as they wanted to during the study, and how frequently would they want to update the clinic on the progress/status of their symptoms. With respect to the Fitbits, we asked if participants looked at their data and about their overall experience using the activity and sleep tracker. Additionally, we shared their average daily sleep and activity data with them. If there were abnormalities, inconsistencies or gaps in their data, we asked participants about it to understand if there were any specific reasons or challenges during the study. Additional questions included asking for context related to clinical recommendations that they reported as not being followed via their surveys and if they talked about their CLH symptoms to other doctors outside of the post-COVID clinic. The interview guide can be found in Appendix B. All interviews were conducted via Microsoft Teams and consisted of the patient participant, a research scientist, and a clinical research nurse. Duration of interviews ranged from 24-69 minutes. Audio of interviews was recorded and transcribed for analysis.

\subsection{Data Analysis} 
We analyzed the collected data using qualitative methods, descriptive statistics, and visualizations. Exploratory data analysis, in the form of descriptive statistics and visualizations, was conducted on survey, Fitbit (activity, sleep) and EHR data. Interviews were analyzed using reflexive thematic analysis~\cite{braun2019reflecting}. We expand upon the analysis for each data stream below.

\subsubsection{Weekly Surveys} 
Weekly surveys were assessed using descriptive statistics. Additionally, we created visualizations to present the changes reported during the study period. Participants were asked if their overall health was worse, the same, or better than the previous week. These were assigned values of -1, 0 and 1 and reported out where appropriate. All data are presented as an average or the last value that was collected as several participants had missing surveys over the duration of the study. Data from the surveys were also used to seed the exit interview, allowing the research team to collect context for missing or irregular data from the weekly surveys. 

\subsubsection{Fitbit} 
Descriptive statistics were calculated for all Fitbit data. The weekly deep sleep duration average was calculated (see Table~\ref{tab:fiitbit_ALL}) due to the critical importance of deep sleep in nervous system recovery, memory consolidation, and immune system reinforcement \cite{Ehrenberg}. Weekly step count averages were used as a proxy for physical activity. Certain aspects of demographics and overall health status were used to compare sleep and activity data to explore if trends were present. We again used visualizations to help better understand trends found from comparing data from different streams within this research. 

\subsubsection{EHR} We conducted retrospective chart reviews \cite{Vassar} to assess notes related to the patients' initial PPCC visit (e.g. demographics, top 5 symptoms, care plan) and the healthcare utilization of patients after their initial PPCC visit. Patient portal messages and phone calls were not included in this assessment. Encounters that did count as healthcare utilization included office visits, emergency room and urgent care visits, hospital admissions, telehealth appointments, and diagnostic procedures/labs. Data collected was compared against patient-reported weekly surveys.

\subsubsection{Interviews} The interviews were transcribed and subjected to reflexive thematic analysis~\cite{braun2019reflecting} wherein the first and second author individually read and open-coded each interview line-by-line and then compared their analyses. Based on patterns in the data and using a combination of open and axial coding~\cite{corbin2014basics}, codes such as \textit{``For self-interest \& self-awareness''}, \textit{``To answer questions \& make comparisons with others''}, and \textit{``To assess general activity levels \& overall health progress''} were clustered and used to formulate higher-levels themes such as \textit{``Reasons for engaging \& interacting with self-tracked health data''}. These themes were used to shape the results.

Although codebooks are not part of reflexive thematic analysis, we still documented codes and initial themes to facilitate coordination across the researcher team, contextualize the different types of data collected from each participant, and present a distribution of the various themes that emerged from this cohort study. 

\subsection{Ethical Considerations of the Research Approach}
While obtaining IRB approval is a critical component in ensuring the safety of research participants, it is not necessarily a comprehensive ethical review of the research methods themselves \cite{Vitak16}. The development of this study went through a rigorous design phase that was approved by the PPCC Clinical Director prior to being submitted to the IRB. This study design goes beyond the traditional HCI field study in that we not only collected Fitbit, interview, and survey data, but also collected data directly from the participants' EHRs. Here we outline the various precautions and actions taken to ensure the safety of our participants and their data. 
\begin{itemize} 
    \item Training/specializations of the research team. For the clinical chart reviews, all team members that touched EHR data went through rigorous EHR training and were credentialed by the hospital system to have access to the EHR. A research nurse on the team with over 20 years of experience oversaw the data validation efforts to ensure that the appropriate data was being abstracted. Team members who conducted the interviews were QPR (Question.Persuade.Refer) trained. This program trains individuals to recognize early warning signs of emotional distress and/or suicidality and connect individuals to the appropriate care \footnote{https://qprinstitute.com/}. The importance of this training will become apparent in Section 4.3. 
    \item Collection and sharing of sensitive data. The informed consent form (ICF) was very explicit about not just what participants would be responsible for (e.g., weekly surveys, interviews, wearing a Fitbit) but also what the researchers would be responsible for (EHR data collection, sharing de-identified/cumulative data back to the clinic). Additionally, a HIPAA authorization was part of the ICF which authorizes the health system to disclose health information in the EHR for the proposed research. The informed consent has been provided as a supplementary document. 
    \item Storing/Treatment of sensitive data. Finally, due to the sensitivities of HIPAA data and the associated penalties if data is exposed \footnote{https://www.ama-assn.org/practice-management/hipaa/hipaa-violations-enforcement}, extra precautions were put in place for the EHR data, including encryption and additional access restrictions. All data storage practices met requirements of the health system.
\end{itemize}

\subsection{Author Positionality}
Although none of the researchers have direct personal experience with CLH issues, they do have close friends and relatives who have experienced CLH symptoms including fatigue, weakness, and brain fog with severity ranging from mild to severe. The research team is interdisciplinary. One team member is the clinical director of the PPCC, several have worked with the PPCC since its inception developing the clinic data registry, and others are affiliated researchers from nursing, HCI, and data science. Several of the non-clinical team members have also shadowed within the clinic to better understand the clinical environment. Thus, this team has unique qualifications to assess the different types of data collected from this research.

\section{Results} 
We present the results from each of the data streams below. We cross reference data where appropriate, highlighting instances where triangulating different data streams provides deeper context and neutralizes potential assumptions related to certain aspects of data related to COVID and CLH.

\subsection{Survey Data}
The weekly surveys collected data on participants' top 5 symptoms, the ranking of their symptoms based on severity, how participants felt about the status of their symptoms compared to the previous week (e.g., worse, same, better), the progress that was made towards completing their clinical recommendations, and their assessment of their overall health relative to the previous week (see Appendix A).

\begin{table*}[ht]
\begin{tabular}{lll}
\textbf{Participant ID} & \textbf{\begin{tabular}[c]{@{}l@{}}Baseline Top 5 Symptoms\\ (ranked from worst to least worst)\end{tabular}} & \textbf{\begin{tabular}[c]{@{}l@{}}Last Survey Top 5 Symptoms\\ (ranked from worst to least worst)\end{tabular}} \\ \hline
P10 & \begin{tabular}[c]{@{}l@{}}weakness, hoarse after talking, \\ coughing, fatigue\end{tabular} & weakness \\ \hline
P11 & \textit{\begin{tabular}[c]{@{}l@{}}memory loss, brain fog, headache,\\ twitches, speech\end{tabular}} & \textit{\begin{tabular}[c]{@{}l@{}}memory loss, brain fog, headache,\\ twitches\end{tabular}} \\ \hline
P12 & \textit{\begin{tabular}[c]{@{}l@{}}fatigue, shortness of breath, weakness,\\ dizziness, brain fog/memory\end{tabular}} & \textit{\begin{tabular}[c]{@{}l@{}}fatigue, shortness of breath, weakness,\\ dizziness\end{tabular}} \\ \hline
P13 & \textit{cough, /congestion} & \textit{cough, congestion} \\ \hline
P14 & \begin{tabular}[c]{@{}l@{}}heaviness in chest, lack of energy,\\ anxiety\end{tabular} & tightness in chest, fatigue \\ \hline
P15 & \textit{\begin{tabular}[c]{@{}l@{}}smell is off, taste, brain fog, sharp\\ pains, sinus congestion\end{tabular}} & \textit{smell is off, taste, brain fog} \\ \hline
P16 & \begin{tabular}[c]{@{}l@{}}nausea, chills, achy, shortness of \\ breath\end{tabular} & headache, general aches/pain \\ \hline
P17 & \begin{tabular}[c]{@{}l@{}}nausea, headache, brain fog, sharp\\ pains, sinus congestion\end{tabular} & \begin{tabular}[c]{@{}l@{}}nausea, headache, diarrhea, general\\ aches/pain\end{tabular} \\ \hline
P19 & \begin{tabular}[c]{@{}l@{}}shortness of breath, rib pain, smell,\\ taste, fatigue\end{tabular} & \begin{tabular}[c]{@{}l@{}}shortness of breath, taste, fatigue,\\ smell\end{tabular} \\ \hline
P20 & \begin{tabular}[c]{@{}l@{}}hand pain, arm pain, fatigue, shortness\\ of breath, headaches/migraines\end{tabular} & \begin{tabular}[c]{@{}l@{}}fatigue, general aches/pain, brain fog,\\ headache\end{tabular} \\ \hline
P21 & \begin{tabular}[c]{@{}l@{}}loss of "real" voice, tightness in chest,\\ lack of concentration, fatigue, \\ headaches\end{tabular} & \begin{tabular}[c]{@{}l@{}}fatigue, headaches, lack of \\ concentration, loss of "real" voice, \\ tightness in chest\end{tabular} \\ \hline
P22 & \begin{tabular}[c]{@{}l@{}}breathing, feel a catch in chest, fatigue,\\ hand tremors, memory issues\end{tabular} & \begin{tabular}[c]{@{}l@{}}shortness of breath, fatigue, tightness in\\ chest, brain fog, leg swelling\end{tabular} \\ \hline
P23 & shortness of breath, fatigue, chest pain & fatigue \\ \hline
P24 & sense of taste, sense of smell, memory & \begin{tabular}[c]{@{}l@{}}smell, taste, memory, brain fog,\\ vertigo\end{tabular} \\ \hline
\end{tabular}%
\caption{Progression of reported symptoms and severity at baseline and last gathered survey. Italics indicates that there was no change in reported symptoms or their severity between baseline and last survey.}
\label{tab:symptoms}
\end{table*}

\begin{table*}[ht]
\begin{tabular}{lcccl}
\textbf{\begin{tabular}[c]{@{}l@{}}Part.\\ ID\end{tabular}} & \multicolumn{1}{l}{\textbf{\begin{tabular}[c]{@{}l@{}}Number of \\ surveys   \\ completed\end{tabular}}} & \multicolumn{1}{l}{\textbf{\begin{tabular}[c]{@{}l@{}}Number of \\ surveys where \\ symptoms\\ changed\end{tabular}}} & \multicolumn{1}{l}{\textbf{\begin{tabular}[c]{@{}l@{}}Number of \\ surveys where \\ severity ranking\\ changed\end{tabular}}} & \textbf{\begin{tabular}[c]{@{}l@{}}Individual \\ symptoms \\ reported to be \\ getting WORSE\\ {[}week(s){]}\end{tabular}} \\ \hline
P10 & 10 & 1 & - & \multicolumn{1}{c}{-} \\ \hline
P11 & 11 & 1 & - & 1 symptom {[}2,3,6{]} \\ \hline
P12 & 12 & - & - & \begin{tabular}[c]{@{}l@{}}1 symptom {[}4,8{]}\\ 2 symptoms {[}5,6,7,9,10{]}\end{tabular} \\ \hline
P13 & 12 & - & - & 1 symptom {[}4,10,11{]} \\ \hline
P14 & 12 & 1 & - & 1 symptom {[}8{]} \\ \hline
P15 & 12 & - & - & 1 symptom {[}4,5,7,8{]} \\ \hline
P16 & 7 & 2 & 5 & \multicolumn{1}{c}{-} \\ \hline
P17 & 12 & 1 & 1 & \begin{tabular}[c]{@{}l@{}}1 symptom {[}7,8,12{]}\\ 2 symptoms {[}9{]}\\ 3 symptoms {[}11{]}\end{tabular} \\ \hline
P19 & 7 & - & 1 & 2 symptoms {[}6{]} \\ \hline
P20 & 11 & 2 & 5 & \begin{tabular}[c]{@{}l@{}}1 symptom {[}1,4{]}\\ 2 symptoms {[}2{]}\\ 3 symptoms {[}3,5,9,11{]}\\ 4 symptoms {[}7,10{]}\\ 5 symptoms {[}8,12{]}\end{tabular} \\ \hline
P21 & 12 & - & 2 & \begin{tabular}[c]{@{}l@{}}1 symptom {[}2{]}\\ 2 symptoms {[}1,3,7,12{]}\\ 3 symptoms {[}4,11{]}\end{tabular} \\ \hline
P22 & 12 & 2 & 3 & 1 symptom {[}12{]} \\ \hline
P23 & 12 & 2 & - & \multicolumn{1}{c}{-} \\ \hline
P24 & 12 & 1 & 2 & 1 symptom {[}1{]} \\ \hline
\end{tabular}
\caption{Symptom changes captured by weekly survey.}
\label{tab:symptom_change}
\end{table*}

\subsubsection{Symptoms} Participants reported a total of 39 unique symptoms. Each participant reported at least 2 and as many as 10 different symptoms over the course of the study. The five most severe symptoms, per the participant rankings, were memory loss, fatigue/weakness, cough, shortness of breath/breathing complications, and loss of taste/smell. These results characterize the heterogeneous nature of CLH in our patient group, with its manifestation widely varying from participant to participant. Table \ref{tab:symptoms} highlights, for each participant, the top 5 symptoms that were reported to the clinic during their initial appointment (baseline) and the top 5 symptoms that were collected on the participant's last survey. Over the course of the study, only one3 of the 14 participants reported no change in their top 5 symptoms or in how they ranked the severity of the symptoms (P11, P13, P15). Within the 12 week study period, 4 of the 14 participants had experienced a new symptom severe enough to be in their top 5 symptoms (P17, P20, P22, P24). Table \ref{tab:symptom_change} highlights the changes in severity ranking and symptoms reported by participants during the study period. 

\begin{figure*}[ht]
    \centering
    \includegraphics[scale=1.0]{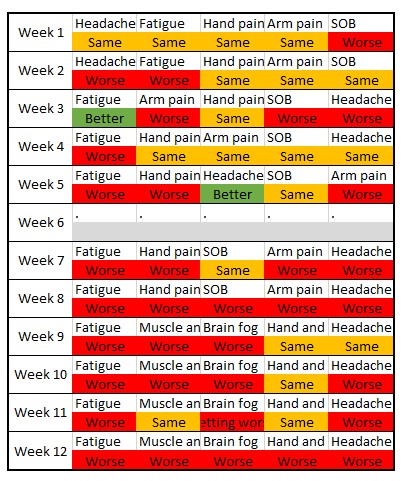}
    \caption{P20 survey results regarding symptom progression with reflections of the status of each symptom for each week for 12 weeks}
    \label{fig:symp_progress_p20}
\end{figure*}

Additionally, in order to develop a more complete picture of the participants' symptom journeys, each week participants were asked to report if their symptoms were getting better, worse, or if they stayed the same, regardless if the symptom was new or if the rankings of their overall severity had changed. Figure \ref{fig:symp_progress_p20} demonstrates P20's data and how complex and dynamic changes in symptoms and severity could be across the 12 week study period. This is further captured in Table \ref{tab:symptom_change} for all participants, depicting how frequently participants reported individual symptoms as getting worse. Only 3 participants (P10, P16, P23) never reported a symptom getting worse over the 12 weeks. For most participants, individual symptoms were getting worse all throughout the study period, with several participants showing steep declines from Week 7 onwards (P12, P17, P20, P21). 

\begin{table*}[ht]
\begin{tabular}{lllllllllllllll}
 & \textbf{P10} & \textbf{P11} & \textbf{P12} & \textbf{P13} & \textbf{P14} & \textbf{P15} & \textbf{P16} & \textbf{P17} & \textbf{P19} & \textbf{P20} & \textbf{P21} & \textbf{P22} & \textbf{P23} & \textbf{P24} \\ \hline
\textbf{Not Started} & \multicolumn{1}{c}{-} & \multicolumn{1}{c}{2} & \multicolumn{1}{c}{7} & \multicolumn{1}{c}{-} & \multicolumn{1}{c}{-} & \multicolumn{1}{c}{1} & \multicolumn{1}{c}{-} & \multicolumn{1}{c}{-} & \multicolumn{1}{c}{5} & \multicolumn{1}{c}{3} & \multicolumn{1}{c}{2} & \multicolumn{1}{c}{2} & \multicolumn{1}{c}{2} & \multicolumn{1}{c}{1} \\ \hline
\textbf{In Progress} & \multicolumn{1}{c}{-} & \multicolumn{1}{c}{-} & \multicolumn{1}{c}{11} & \multicolumn{1}{c}{6} & \multicolumn{1}{c}{4} & \multicolumn{1}{c}{5} & \multicolumn{1}{c}{4} & \multicolumn{1}{c}{9} & \multicolumn{1}{c}{4} & \multicolumn{1}{c}{5} & \multicolumn{1}{c}{4} & \multicolumn{1}{c}{9} & \multicolumn{1}{c}{7} & \multicolumn{1}{c}{4} \\ \hline
\textbf{Completed} & \multicolumn{1}{c}{15} & \multicolumn{1}{c}{2} & \multicolumn{1}{c}{3} & \multicolumn{1}{c}{6} & \multicolumn{1}{c}{9} & \multicolumn{1}{c}{3} & \multicolumn{1}{c}{5} & \multicolumn{1}{c}{4} & \multicolumn{1}{c}{6} & \multicolumn{1}{c}{4} & \multicolumn{1}{c}{1} & \multicolumn{1}{c}{7} & \multicolumn{1}{c}{7} & \multicolumn{1}{c}{4} \\ \hline
\textbf{Non-Adherence} & \multicolumn{1}{c}{1} & \multicolumn{1}{c}{12} & \multicolumn{1}{c}{-} & \multicolumn{1}{c}{1} & \multicolumn{1}{c}{4} & \multicolumn{1}{c}{2} & \multicolumn{1}{c}{8} & \multicolumn{1}{c}{1} & \multicolumn{1}{c}{3} & \multicolumn{1}{c}{6} & \multicolumn{1}{c}{6} & \multicolumn{1}{c}{-} & \multicolumn{1}{c}{3} & \multicolumn{1}{c}{4} \\
\multicolumn{1}{r}{\textit{Clinical}} & \multicolumn{1}{c}{\textit{1}} & \multicolumn{1}{c}{\textit{7}} & \multicolumn{1}{c}{\textit{-}} & \multicolumn{1}{c}{\textit{1}} & \multicolumn{1}{c}{\textit{4}} & \multicolumn{1}{c}{\textit{2}} & \multicolumn{1}{c}{\textit{3}} & \multicolumn{1}{c}{\textit{1}} & \multicolumn{1}{c}{\textit{3}} & \multicolumn{1}{c}{\textit{6}} & \multicolumn{1}{c}{\textit{5}} & \multicolumn{1}{c}{\textit{-}} & \multicolumn{1}{c}{\textit{3}} & \multicolumn{1}{c}{\textit{4}} \\
\multicolumn{1}{r}{\textit{Lifestyle}} & \multicolumn{1}{c}{\textit{0}} & \multicolumn{1}{c}{\textit{5}} & \multicolumn{1}{c}{\textit{-}} & \multicolumn{1}{c}{\textit{0}} & \multicolumn{1}{c}{\textit{0}} & \multicolumn{1}{c}{\textit{0}} & \multicolumn{1}{c}{\textit{5}} & \multicolumn{1}{c}{\textit{0}} & \multicolumn{1}{c}{\textit{0}} & \multicolumn{1}{c}{\textit{0}} & \multicolumn{1}{c}{\textit{1}} & \multicolumn{1}{c}{\textit{-}} & \multicolumn{1}{c}{\textit{0}} & \multicolumn{1}{c}{\textit{0}} \\ \hline
\end{tabular}
\caption{Final self-reported adherence to individual care plan recommendations. Reported numbers represent the total individual care plan recommendations for each category}
\label{tab:careplan_adherence}
\end{table*}

\subsubsection{Care Plan}
Participants were asked each week to report on the status of individual components of their PPCC care plans. They could report that they had not started making progress, were making progress, had completed the recommendation, were not planning on addressing the recommendation or that the recommendation did not apply to them. Each participant was given a unique care plan tailored to their specific needs in addition to the 8 recommendations that were common across all participants (5 lifestyle recommendations including hydration/diet, exercise, sleep, weight management, stress management; medication adherence; COVID vaccination; participation in the PPCC support group). The average number of unique, patient-specific recommendations was 8 (SD=2.8), and when taking into consideration all of the common and unique recommendations, the average number of recommendations was 16.1 (SD=2.8). Table \ref{tab:careplan_adherence} highlights the final breakdown of recommendation progress for each participant. 

We define non-adherence as participants reporting they did not plan on following a given recommendation or felt the specific recommendation did not apply to them. Only 2 participants did not report a non-adherence measure (P12, P22). On average, 23\% of the care plan recommendations were reported as a non-adherence, with the maximum being P11 with 75\% non-adherence. Adhering to clinical recommendations and adhering to daily lifestyle habit changes are arguably different processes and require different levels of motivation and planning. Table \ref{tab:careplan_adherence} depicts the participants' reporting of the progress they had made on their individual care plan elements. We found that almost all participants were highly polarized, either choosing to not adhere to clinical recommendations or to lifestyle-based recommendations. In the interviews, participants were asked about every care plan element where they had indicated non-compliance. Section 5.3.1 goes into more detail about the barriers related to compliance, most reasons were due to personal preference. Examples of this include "just not wanting to do it", "don't have time", or believe it to be "uncomfortable". Other rationale included financial considerations and symptoms resolving/not feeling like further attention is needed. Fear that additional issues would be caused by care plan activities was the other key reason provided.\\

\noindent{\textbf{Validating Care Plan Adherence Reports with EHR Data}: 
In relation to the survey data collected on PPCC care plan recommendation adherence, we reviewed EHR data for the six months following the participants' initial PPCC visit. Table \ref{tab:EHR_follow} highlights what was uncovered, broken down by type of clinical recommendation. Almost half (6/14) of the participants had conflicting information between the self-reports on the weekly surveys and what was validated through the manual chart review for each participant. The most common conflict was associated with follow-up on orthostatic blood pressure and heart rate monitoring and rehabilitation therapies. While we reviewed the patient EHRs for these aspects, they are also highly probable to not be reported with the EHR, especially if rehabilitation support is found with other local health providers that are not part of the health system or something that can be done at home with specialized equipment and care support like orthostatic blood pressure and heart rate monitoring. }The most frequent type of testing that was completed by participants was cardiovascular testing with all but 1 participant (P20) completing such recommendations. Neurological tests received the least amount of adherence with 3 (P11, P12, P20) of the 5 participants who were recommended testing not having any indication in the EHR that neurological testing had been scheduled and/or completed.

\begin{table*}[ht]
\begin{tabular}{lcccccccc}
\textbf{\begin{tabular}[c]{@{}l@{}}Part. \\ ID\end{tabular}} & \multicolumn{1}{l}{\textbf{\begin{tabular}[c]{@{}l@{}}COVID\\ Vacc.\end{tabular}}} & \multicolumn{1}{l}{\textbf{\begin{tabular}[c]{@{}l@{}}Ortho\\ BP/HR\end{tabular}}} & \multicolumn{1}{l}{\textbf{Labs}} & \multicolumn{1}{l}{\textbf{\begin{tabular}[c]{@{}l@{}}Specialist \\ Appt.\end{tabular}}} & \multicolumn{1}{l}{\textbf{\begin{tabular}[c]{@{}l@{}}PCP\\ Follow\\ Up\end{tabular}}} & \multicolumn{1}{l}{\textbf{\begin{tabular}[c]{@{}l@{}}PPCC\\ Follow\\ Up\end{tabular}}} & \multicolumn{1}{l}{\textbf{Testing}} & \multicolumn{1}{l}{\textbf{\begin{tabular}[c]{@{}l@{}}Rehab\\ Therapies\end{tabular}}} \\ \hline
P10 & Yes & Yes & Yes & \begin{tabular}[c]{@{}c@{}}Yes (2)\\ No (1)\end{tabular} & Yes & N/A & N/A & N/A \\ \hline
P11 & Yes & No & No* & N/A & N/A & No & \begin{tabular}[c]{@{}c@{}}Yes (1)\\ No (2)\end{tabular} & No \\ \hline
P12 & Yes & No & No & Yes (1) & No & N/A & \begin{tabular}[c]{@{}c@{}}Yes (1)\\ No (2)\end{tabular} & No \\ \hline
P13 & Yes & No & N/A & Yes (1) & Yes & N/A & Yes (2) & N/A \\ \hline
P14 & No & No & Yes & N/A & Yes & N/A & Yes (1) & No* \\ \hline
P15 & No & No* & Yes & No (2) & N/A & N/A & N/A & N/A \\ \hline
P16 & Yes & No & No & N/A & N/A & N/A & N/A & N/A \\ \hline
P17 & No & No & Yes & No (1) & N/A & N/A & Yes (3) & Yes \\ \hline
P19 & Yes & No & Yes & Yes (1) & Yes & N/A & N/A & Yes \\ \hline
P20 & Yes & No* & N/A & No (2) & N/A & N/A & No (4) & No \\ \hline
P21 & No & No* & N/A & Yes (1) & N/A & No & N/A & No \\ \hline
P22 & No & No & Yes & No (1) & N/A & N/A & Yes (2) & Yes \\ \hline
P23 & No & No* & Yes & N/A & N/A & N/A & \begin{tabular}[c]{@{}c@{}}Yes (2)\\ No (1)\end{tabular} & N/A \\ \hline
P24 & Yes & No & Yes & N/A & N/A & N/A & N/A & N/A \\ \hline
\end{tabular}
\caption{Validating progression of care plan activities reported in surveys against EHR data.
\textit{* indicates where there was a different response in the participant's survey}}
\label{tab:EHR_follow}
\end{table*}

\subsubsection{Overall health progression}
\begin{figure*}[ht]
    \centering
    \includegraphics[scale=1.0]{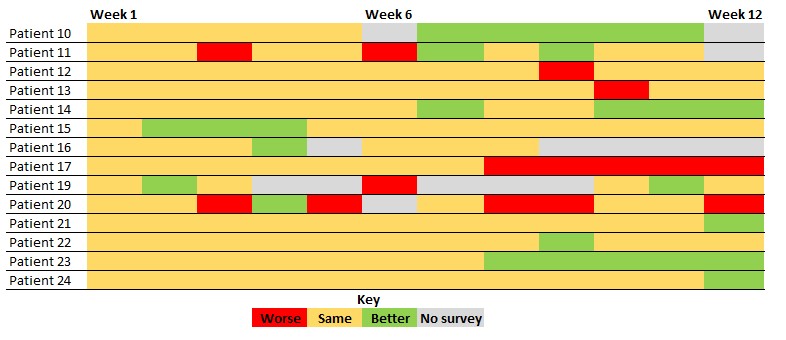}
    \caption{Progression of overall health assessment}
    \label{fig:symp_progress}
\end{figure*}

Only 3 participants (P12, P13, P17) reported no positive progression of their overall health. Figure \ref{fig:symp_progress} highlights the reported changes each week. Although some participants showed high variation in their overall health (P11, P19, P20), the most common trend was long periods of no change with signs of getting better towards the end of the 12 week study period. When comparing participants who had completed most of their care plan (see Table \ref{tab:careplan_adherence}) to reports of overall health, the 2 most adherent participants (P10, P14) reported their overall health as getting better. On the other hand, the 2 least adherent participants (P11, P20) show very mixed responses and lack of stable health, wavering between ``better'', ``worse'', and ``about the same.'' 

Table \ref{tab:fiitbit_ALL} highlights many components, including a summative "overall health" score. We applied numeric values to the three options: worse (-1), about the same (0), and better (+1) from the weekly surveys. When comparing the changes each week, these follow a naturalistic distribution where 1/2 were in a mid-range, 1/3 were leaning towards consistent better health and 1/6 were leaning towards consistent worsening health.

\subsection{Fitbit Data}
Tables \ref{tab:fiitbit_ALL} and \ref{tab:fitbit} show individual and summative breakdowns of the Fitbit data. We see that across the board, sleep values and activity of the study participants are all below the ranges suggested by leading clinical entities \cite{mayo}. We present the activity and sleep data as a way to compare within the group and not comparisons to community-held norms. Data for 3 participants are not reported. P11 and P15 did not consistently use the Fitbit during the study period and P19 was unable to effectively use the Fitbit due to extreme hand tremors brought on by her CLH. P19 reported in the interview that she could not get the magnetic charger to stay on due to this issue, became frustrated, and stopped wearing the device all together. 

\begin{table*}[ht]
\begin{tabular}{lccccccc}
\textbf{\begin{tabular}[c]{@{}l@{}}Part.\\ ID\end{tabular}} & \multicolumn{1}{l}{\textbf{\begin{tabular}[c]{@{}l@{}}Age\\ Range\end{tabular}}} & \multicolumn{1}{l}{\textbf{\begin{tabular}[c]{@{}l@{}}Overall\\ Health\end{tabular}}} & \multicolumn{1}{l}{\textbf{\begin{tabular}[c]{@{}l@{}}Avg. Steps\\ (daily)\end{tabular}}} & \multicolumn{1}{l}{\textbf{Median}} & \multicolumn{1}{l}{\textbf{\begin{tabular}[c]{@{}l@{}}Avg. Sleep \\ (daily/hours)\end{tabular}}} & \multicolumn{1}{l}{\textbf{Median}} & \multicolumn{1}{l}{\textbf{\begin{tabular}[c]{@{}l@{}}Sleep (deep)\end{tabular}}} \\ \hline
P10 & 71-80 & 0.5 & 6548 & 6848 & 7.0 & 8.1 & 14.0\% \\ \hline
P12 & 51-60 & -0.1 & 5501 & 5774 & 6.2 & 6.7 & 16.3\% \\ \hline
P13 & 51-60 & -0.1 & 8382 & 7915 & 3.6 & 3.9 & 13.5\% \\ \hline
P14 & 71-80 & 0.4 & 2471 & 2424 & 4.8 & 5.3 & 8.5\% \\ \hline
P16 & 41-50 & 0.2 & 4873 & 4847 & 3.4 & 4.5 & 12.6\% \\ \hline
P17 & 51-60 & -0.4 & 1311 & 985 & 7.9 & 7.9 & 12.1\% \\ \hline
P20 & 31-40 & -0.4 & 7401 & 8515 & 4.2 & 4.7 & 14.8\% \\ \hline
P21 & 31-40 & 0.1 & 6674 & 6867 & 4.8 & 6.8 & 15.6\% \\ \hline
P22 & 51-60 & 0.0 & 5728 & 5816 & 4.9 & 5.3 & 11.6\% \\ \hline
P23 & 31-40 & 0.5 & 13862 & 13920 & 4.0 & 4.1 & 20.5\% \\ \hline
P24 & 41-50 & 0.1 & 7518 & 7085 & 5.7 & 6.3 & 17.8\% \\ \hline
\end{tabular}
\caption{Fitbit data. Patients P11, P15, P19 removed for lack of data. Overall health defined by weekly surveys.}
\label{tab:fiitbit_ALL}
\end{table*}

\begin{table*}[ht]
\begin{tabular}{llccccc}
 & \textbf{Variable} & \multicolumn{1}{l}{\textbf{Mean}} & \multicolumn{1}{l}{\textbf{St.Dev}} & \multicolumn{1}{l}{\textbf{Min}} & \multicolumn{1}{l}{\textbf{Median}} & \multicolumn{1}{l}{\textbf{Max}} \\ \hline
Sleep & Light (min) & 256.7 & 80.0 & 79 & 253 & 613 \\
 & REM (min) & 75.4 & 37.9 & 0 & 71 & 225 \\
 & Deep (min) & 57.2 & 27.5 & 0 & 54 & 182 \\
 & Awake (min) & 56.4 & 23.2 & 3 & 52.5 & 146 \\
 & \# Times Awake & 20.3 & 7.7 & 2 & 19 & 71 \\ \hline
Activity & Step Count & \multicolumn{1}{c}{5,863.4} & \multicolumn{1}{c}{4,408.5} & \multicolumn{1}{c}{27} & \multicolumn{1}{c}{5,165.5} & \multicolumn{1}{c}{28,965} \\
 & Light (min) & \multicolumn{1}{c}{214.0} & \multicolumn{1}{c}{107.5} & \multicolumn{1}{c}{3} & \multicolumn{1}{c}{213} & \multicolumn{1}{c}{608} \\
 & Moderate/Intense (min) & \multicolumn{1}{c}{40.8} & \multicolumn{1}{c}{85.1} & \multicolumn{1}{c}{0} & \multicolumn{1}{c}{0} & \multicolumn{1}{c}{652}
\end{tabular}
\caption{Summary of Fitbit Data}
\label{tab:fitbit}
\end{table*}

\subsubsection{Sleep} On average, participants slept 5.1 hours per night. The average percentage of deep sleep was 14.3\%. These values are well under the amounts recommended by the CDC (7-9 hours per night; 20\% of total sleep) \cite{national_heart}. Table \ref{tab:fiitbit_ALL} highlights the averages for each participant. Three of the participants (P10, P16, P21) have median sleep values much larger than the means, showing the high volatility in their daily recordings. 

\begin{figure*}[ht]
    \centering
    \includegraphics[scale=0.60]{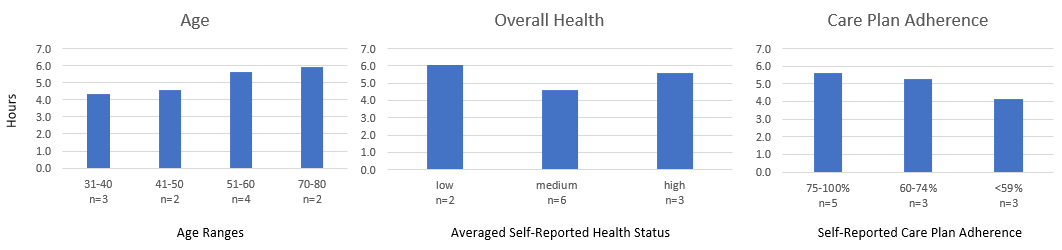}
    \caption{Breakdown of sleep outcomes per key indicators}
    \label{fig:fitbit_sleepall}
\end{figure*}

We examined sleep data as it related to key indicators of age, overall health status, and care plan adherence as reported through the surveys (see Figure \ref{fig:fitbit_sleepall}). The age groupings highlight the younger the participant, the lower the levels of overall sleep, but higher percentages of deep sleep. This trend held consistent through each of the 4 age ranges reported. When we looked at sleep values as it related to the self-reported overall health ranking, average time asleep was not found to have a consistent trend. However, those who reported the least overall positive health had the lowest percentage of sleep that was deep and those that reported the highest had the most. This is congruent with research findings that higher levels of deep sleep are correlated with lower levels of inflammation \cite{Mills07}. Finally, while a positive trend was found between the amount of sleep and the adherence to the care plan, no consistent trend was found between percentage of deep sleep and this attribute.

\subsubsection{Activity} Participants' average number of daily steps was 5,863, which is well under the general 10,000 daily steps recommendation. The average duration of light activity over a day was under 3.6 hours, and moderate to intense activity averaged around 40 min per day, albeit skewed by one particularly active patient (P23). Tables \ref{tab:fiitbit_ALL} and \ref{tab:fitbit} highlight this breakdown. During the interviews it was ascertained that P23 was the only participant who worked a full-time labor-based job. They shared the volatility of their captured activity, needing periods of intense recuperation after highly active days because of their CLH condition. P23's  minimum recorded activity was 1,781 steps versus the maximum of 28,965 steps with a standard deviation of 6,808 steps (the largest individual SD of any participant in the study).

\begin{figure*}[ht]
    \centering
    \includegraphics[scale=0.55]{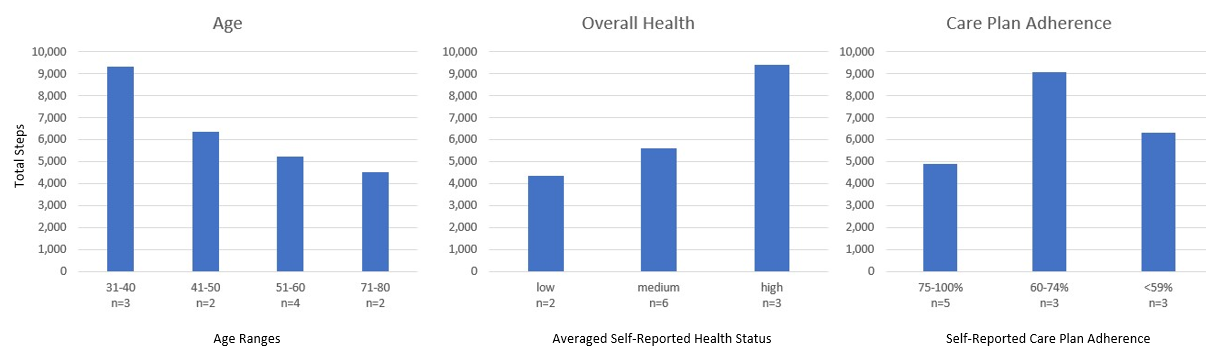}
    \caption{Breakdown of activity outcomes per key indicators}
    \label{fig:fitbit_stepsall}
\end{figure*}

Similar to the sleep data, we looked at activity data through the lens of age, overall health status, and care plan adherence (see Figure \ref{fig:fitbit_stepsall}). The younger the participant, the levels of activity were higher with each increasing age range seeing less and average activity. When looking at the reported health status, participants with lower health status had lower levels of activity. An interesting finding was that those more adherent with their care plan recommendations, which includes physical activity, averaged almost half of the average daily steps as the group with less adherence. One reflection for this could be that while younger patients are more active, they are less adherent to care plan recommendations. However, this sample was too small to assert this as anything other than an observation.

\subsection{Interview Data}
\subsubsection{Barriers to Compliance}
Participants discussed various barriers they faced in adhering to their PPCC care plan, with cost and insurance issues being a common reason for non-compliance. Several participants discussed the actual costs associated with the PPCC healthcare orders and recommendations. For example, one of P11's care plan items was for her to have a specific lab drawn, but she noted challenges with her insurance company in that they were \textit{``just being jerks about things...I'm only on disability and what I earn from my [side job] which isn't a ton, so it is difficult financially to do what I need for my health.''} P11 had complex needs that were exacerbated by cost-related and insurance coverage issues. She shared that a key reason for non-adherence to her CLH care plan was her insurance \textit{``not wanting to cover''} certain tests and appointments, which were \textit{``too expensive''} for her to cover on her own.

Besides healthcare costs, participants also struggled with disharmony between strongly recommended healthcare directives and their lived reality. For instance, the general nature and stresses of work were often too much for CLH patients. They were commonly recommended by providers to reduce stress and hours worked. However, for many participants who had fixed incomes or were in tight financial situations, it was not possible to abide by this recommendation. P12 described the realities of their job in which they were \textit{``working more hours than the doctor wants me [P12] to [because] they don't want me to work at all [due to the] brain damage [resulting from COVID].''} However, they had to \textit{``force''} themselves to work as  \textit{``the bills still keep coming in.''} Similarly, P20 talked about their manual labor job as \textit{``pretty much all or nothing.''} She mentioned her primary doctor telling her to \textit{``lighten the load, you need to work less''} but that not being possible for her: \textit{``I said good luck. Like, you can write me a note all day long, but it's not gonna help me out any.''} 

Another barrier to compliance with the PPCC care plan, or even compliance with research study requirements, was other ongoing participant health issues. For example, P19 discussed that adherence to her care plan was \textit{``definitely affected by multiple hospitalizations''} she had right after her clinic visit. She could not \textit{``get a lot done that [she] needed to get done''} but that she had tried in earnest to \textit{``do what she could and worked on it a little bit''} when she could. This was demonstrated when she answered a follow-up call from the research team while hospitalized. Additionally, we saw participants putting the health needs of their family members before themselves. For example, when asked about her care plan progress, P11 stated:

\begin{quote}
    \textit{I haven't gotten my MRI done. [Family members passed away and were ill] ... so like all of this stuff happened. So it's like everything like that kind of stuff got put on the back burner, taking care of myself while I was dealing with all of my family issues.} (P11)
\end{quote}

Similarly, P20 noted prioritizing ongoing medical issues of their spouse due to their severity and dual cost burden:
\begin{quote}
    \textit{We had just come off of last fall [when my partner was ill and hospitalized]...those bills started rolling in. I see this list of things recommended for me to get done [and the insurance company] weren't wanting to cover them...so those are definitely on the back burner [for now].} (P20)
\end{quote}

Thus, patient care plan compliance was impacted by other life, family, and cost priorities, often leading to them de-prioritizing their own healthcare needs.

\subsubsection{Uncertainty \& New Normals}
Participants described a great deal of uncertainty related to the medical complexities of CLH. Participants discussed frustration in their inability to detect if an issue was related to CLH or something different. For example, P20 talked about uncertainty when she experienced difficulty breathing:
\begin{quote}
    \textit{...like right now because my basement isn't watertight. And when I have to go do laundry, like I can tell that my breathing gets really affected. Just being down there for 5 or 10 minutes.} (P20)
\end{quote}

 This uncertainty was also related to unexpected flare ups of current symptoms, such as when P13 \textit{``ran out of the [medication] somewhere in the middle of [his] vacation''} and was unsure if he had \textit{``caught something being around everybody on vacation''} or if it was a returning cough related to CLH. Participants also discussed their struggles when traditional or lifestyle methods of recovery did not work. For example, despite following PPCC recommendations for exercise and taking regular vitamins, P10 described getting \textit{``tired fairly easily''} as his \textit{``biggest problem.''}

Moreover, the idea of accepting and living a \textit{new normal} emerged as an unprovoked recurrent theme. P22 discussed the genesis of this new reality which emerged while initially hospitalized for her COVID infection. She shared:

\begin{quote}
    \textit{When I was in the hospital with COVID, I thought, OK, I'm gonna go home. I'm just gonna recuperate a couple weeks and it's gonna be done. You know, I'm gonna go back to normal. It's probably never going to happen, you know, especially now with the scar tissue in the lungs and the problems that that's causing. So it's like, OK, let's find a new normal and all of these steps are finding that new normal and it's kind of helping with that.} (P22)
\end{quote}

Participants also described resignation to the uncertainty associated with their new normal. P12 stated:
\begin{quote}
    \textit{I guess one I didn't think to do it because this is just gotten so almost a new normal for me that if there's a change like that, you know it's like, OK, I've had one hell of a day, you know that type of thing.} (P12)
\end{quote}
 
 P15 also shared how she is \textit{``just going with the flow''} and if CLH symptoms come back, then it is just her new normal. This new normal also extended to physical attributes and new accommodations, like P14 who now cannot walk much due to fatigue and now uses a wheelchair which impacts her ability to participate in some activities. This highlights the level of uncertainty consistent across different participants.

Family, friends, and healthcare providers of participants also struggled with this new normal. Often people would ask questions about participants' ongoing health issues. For example, P20 shared that \textit{``people were just kind of like -- Why? Why aren't you doing your work? Why are you so tired?''} Moreover, people would rationalize their questioning and compare the participant's experiences with other people's experiences with CLH: \textit{``[people would comment] You know it wasn't that bad. I heard so. And so my cousin, my neighbor had it and they were fine after a week.''} (P20). It was challenging when this lack of understanding came from healthcare professionals. P17 mentioned that some nurses she knew had \textit{``never even heard of long haul COVID''} which caused a \textit{``breakdown''} in the system as she had to tell them to \textit{``go look it up because it's a thing you need to go do some research.''}. Similarly, P14 shared a stressful experience wherein their primary care provider (PCP) refused to tend to their CLH-related complications:
\begin{quote}
    \textit{My PCP did not want to see me at all. I was in emergency four different times. And the reason being is I was taken to the emergency first time by ambulance. And they said no. Tomorrow you need to see your doctor [and I] would call the doctor's office and he said absolutely not. Go back to ER. So we went back and forth quite a bit and at the end of the third visit, he finally ordered a cat scan of my lungs...I'm currently looking for a new doctor.} (P14)
\end{quote}

Participants also shared how the uncertainty of CLH manifested itself in mental health concerns. A few participants who had mental health issues prior to having COVID detailed how CLH exacerbated their condition(s). For example, P11 spoke of mental health related complications
\begin{quote}
    \textit{I suffer from bipolar disorder, so that doesn't help either because you know not getting the sleep I need and then with the brain fog you know, kind of you feel so stupid and sends you kind of into a depressed state even more than before.} (P11)
\end{quote}

We also found that two participants (P17, P9) were hospitalized for mental health issues during the study period with another mentioning having had suicidal thoughts (P12). P17 shared that stress due to CLH and related uncertainties led her to being hospitalized for suicidal ideation: 
\begin{quote}
    \textit{I was going to attempt suicide... I didn't ultimately do it. I was threatening the night before. I was gonna take some pills and I didn't end up doing it. And I told [my son who called my psychiatrist] in the morning. I thought I almost took pills last night.} (P17)
\end{quote}

However, none of these three participants ever mentioned depression, anxiety or other mental health issues in their top 5 symptoms. None of these participants were on anxiety or depression medications, had a PPCC visit diagnosis of mental health issues, or had a mental health issue on their EHR problem list. 17 had only one indicator, a mental illness in their EHR problem list. However, the date of when this was last verified was unclear. 

Finally, related to their participation in the study, several participants discussed how their symptoms impaired their adherence to the research protocol. P12 and P22 had difficulties remembering to take the surveys if they did not do it right when they received their email and attributed this to their brain fog issues. Hospitalizations were also an issue with respect to completing weekly surveys or wearing their Fitbit (P11, P19). However, when reminded or nudged, most participants appreciated and complied with what was requested. 

\subsubsection{Engaging with Fitbit \& Other Tracked Data}
Participants engaged with Fitbit data primarily for their own self-awareness. P10 shared that he liked his Fitbit \textit{``for the fact [that] I could track myself and make sure I could tell when I was sitting on my rear [too much].''} P13 originally was less pleased with having to wear the device, but became intrigued by the data being generated leading him to profess that \textit{``I want it now, I want to look at it [Fitbit collected health tracking data] and I'll put the damn thing back on so I can do so.''}. Some participants were less engaged with the technology. For example, P11 described checking their tracked Fitbit data intermittently, \textit{``sometimes it would be every day and then other times...it would be once a week... I would get in and check on it, especially if I was really active that day.''} Fitbit visualizations provided participants the ability to make connections between how they were feeling as it related to their sleep and activity. For example, P22 summarized this sentiment as:
\begin{quote}
    \textit{The nice thing is you can see where you started, where you were at midpoint and then this is where you are today. Letting me know how I progressed was kind of nice. The post COVID has affected sleeping and stress management so all those things that you can see on the Fitbit were perfect.} (P22)
\end{quote}

All participants who reported engaging with their Fitbit data, shared a feeling of appreciation for the immediate feedback it provided them. For some participants, it offered a synthesis of other technologies, allowing them to review data for multiple aspects of their health in real time. P22 shared, \textit{``I love that it monitors my oxygen for me because I had an external pulse ox that I was using. Now I can just look at the Fitbit wherever I am and check my heart rate. Then I can see, OK, I need to sit down because I am doing too much or that I'm good, that I can keep going...because I generally don't know [until it's too late].}

Only one participant discussed how the data was used in collaboration with a clinician. Several mentioned to a physician that they were tracking their data as part of a study, but only P22 described their discussions with a physical therapist about using a Fitbit to track activity and the need to better understand when not to push their body - a key element to early post-COVID recovery. 

When participants were asked if they would continue to use their Fitbit after the completion of the study, responses were mixed. Nine participants responded that they would continue tracking their activity. For these 9, the general consensus was that the Fitbit data made them more aware of their health status at any given time (P11). P12 liked the Fitbit so much they wanted to upgrade to a more sophisticated model. On the other hand, P15 mentioned liking the multi-tasking of the Apple watch better, so said she will still track her health but using another device. Some participants were less committed, responding that while they might not wear a health tracking device like the Fitbit everyday, they had worn the Fitbit sporadically since the conclusion of the study (P16). For those who would not continue to track their activity after the study concluded,  major issues contributing  to this decision included problems syncing the Fitbit with their phone (P10, P11) and issues with the device form factor and charging (e.g., P19 experienced Post-COVID hand tremors which made it difficult for her to wear the Fitbit and connect it to a charger). Others mentioned not normally wearing a watch, so it felt unnatural to continue wearing the device (P10). 

Three participants (P11, P14, P23) were also interested in continuing to track their symptoms and progress on their care plans. For example, P11 was enthusiastic about continuing their symptom tracking in the future because \textit{``it has made me more aware, which is so great!''} P13 mentioned that not only was he tracking symptoms, but had gotten his wife involved as well since she was also having lingering issues from a COVID infection.

\subsubsection{Impacts of Study on Participants} A common theme that came up during the interviews was the impact that just being in the research study had on the participant. This was not probed on, but came out in the course of discussion. P10 had never had a gym membership before, but because of the combination of wearing the Fitbit and the recommendation to increase exercise, he signed-up at the local YMCA and created a self-paced plan to start gradually moving his body. He shared that: \textit{``I needed something to push... to get moving and I think this did help me.''} He went on to state that: \textit{``When you were checking on me every week, it gave me something to look forward to... To see how I am doing. You know the update, ... I used [it] as an encouragement.''} Other participants discussed a sense of isolation within their CLH and that interacting with the researchers and receiving weekly correspondence about the survey helped with that. For example, P11 stated: \textit{``I am so glad you guys are doing this because a lot of research, you know it needs to be done and having you all do it, it's just it's nice because it does make me feel less alone... I really, I really appreciate the support and everything.''} Others discussed how important it was for them to participate because they wanted to help others with these issues. P12 highlights this sentiment: \textit{``I've always tried to help people. I mean [if] you can learn something from me that helps somebody else down the road... it's worth it.''}.

\section{Discussion}
In each section, we discuss various facets of our findings, how they relate back to or challenge our current knowledge about CLH and how data can drive design. We provide policy and design provocations in addition to aspects that we believe are ripe for future work.

\subsection{New Normals \& the Use of Technology in Defining CLH}
All participants discussed challenges and complexities related to living with CLH and defining their ``new normal.'' The challenges associated with an illness lacking a standardized means of diagnosis creates uncertainty and anxiety about having a health condition where there are more questions than answers. As seen in the results, several participants (e.g., P20, P13) talked about not knowing if their symptom was new and related to CLH or just a part of growing older or something else altogether. Being treated at the PPCC was seen as a major validation of their issues being "real." Our findings on everyday burdens and issues arising from having a lesser known chronic illness validate similar findings from other HCI studies on chronic illness \cite{MacLeod}. As previous work has shown, communicating about chronic illness can be challenging because people are often self-conscious of their conditions \cite{Pang}. However, we do know that sharing personal health data to build a community has proven to support people with a variety of health conditions from cancer \cite{Rosenberg} to rare diseases \cite{MacLeod}. In Bardzell et. al., the idea of creating decentralized networks to support those facing the uncertainty associated with menopause and how this helped inform women's new normal is particularly relevant \cite{Bardzell}. A recent study of men experiencing fertility issues explored aspects of creating support networks online and using this community to make sense of this new reality \cite{Patel}. By providing a safe space to socially construct the normal, the online community serves as a platform that can bridge the gaps between understanding and acceptance \cite{Shelagh}.

Through our interviews, we uncovered that people were routinely discredited by clinical and non-clinical individuals. Thus, a part of the new normal, was sometimes feeling even more isolated in their disease. Technology can help address or alleviate isolation by connecting people to others for finding support \cite{Day,Pater23}. Future technologies should leverage health data and patient-reported needs to connect people in a more meaningful way rather than just creating unstructured and/or omnibus environments for people to seek out support. Can we use sophisticated modeling or AI approaches to lessen the burden of people finding online support? If this is technically and socially feasible, what are the ethical implications of creating and supporting this type of technology? General social media platforms and social technologies could be more proactive in engaging users directly on their desires to connect to others based on search or posting parameters, especially as it relates to health. These platforms are often black boxes with respect to the metadata around platform utilization and how algorithms connect them to people and content. Thus, they are uniquely positioned to play a more instrumental, proactive role in deepening connections within their platforms. 

Visibility is also a major component in defining the new normal. Some participants dealt with physically visible aspects that had to be accepted, like now relying on a wheelchair (P14) while most others dealt with more invisible aspects like fatigue and mental health concerns which, at times, were exacerbated by a lack of understanding and/or support by other people (see Section 5.4). Whether or not people disclose a chronic illness is often related to the stigma associated with both the visible and invisible aspects of the chronic health issue \cite{Joachim}, and for CLH patients, this can be further exacerbated by the nascent nature of the illness itself. However, most participants displayed resilience in this evolving situation, referring to the need to accept new limitations while at the same time continuing to look for treatments and answers to ensure this new normal was not just giving in to their current medical state.

The technology used in this study was also connected to defining new normals. P10 talked about completely changing his orientation around exercise initially to be compliant with the study but through that compliance, created a new and meaningful health behavior. This type of behavior change can be challenging, but technology has been shown to help some individuals get to that point \cite{Gowin}. Although some participants used technology to help develop new health behaviors, others used the device as a part of their outward identity. The way our participants used their Fitbits and talked about this in their interviews supports the concept of ``lived informatics'' and how personal tracking devices and the data they produce are integrated into a person's daily life \cite{Rooksby}. Regardless of the driver of use, technology has a critical role to play in defining the new normal of patients. Because of the various presentations and severity of CLH, technology developed for CLH patients should be developed using a scoping user-centered design approach to ensure that it meets the needs within this community. Using opportunistic design approaches will not suffice, thus deep partnerships with Post-COVID Clinics or community groups focused on CLH will be paramount for designing technology that addresses core clinical needs of the CLH patient and their providers. 

Future technologies designed to support the CLH journey will have to be agile and understand not only visible, but the invisible needs that are associated with supporting this emerging illness. Early stage design research is needed to define and prioritize needs triggered by "new normals" \cite{Orhun}. The mental health challenges presented in this study embodies this idea of invisible needs. Emerging technologies in the telehealth and AI space could provide the "at elbow support" needed for newly diagnosed CLH patients (like those in this study) to provide proactive mental health support. Platforms like Woebot \footnote{https://woebothealth.com/} and Youper \footnote{https://www.youper.ai/} provide digital therapeutics to clients through smart AI and research-backed interventions. Future work is needed to asses the efficacy of integrating these emerging technologies into Post-COVID clinical care.

\subsection{Data Complexities \& Implications for Research Design}
In the study, the idea of complexity was scattered throughout the various data streams. In the surveys, we saw that some participants would report they were making progress towards a recommendation and then the next week they would report they had not yet started the same recommendation. For people with complex cognitive issues such as memory loss or brain fog, traditional surveys may not be the most appropriate tool, and using cognitive interviewing or other methods might yield more reliable data \cite{Gordon}. However, we found this to be problematic as well, with several participants needing to stop during the interviews and be reminded of the question they were just asked. During the study, we sent up to three reminders each week if participants had not taken their surveys. Participants appreciated the reminders as they meant to take the survey, but their brain fog issues caused them to easily forget (see Sections 5.4.1 \& 5.4.4 ). This cognitive issue also impacted adherence to wearing their Fitbits. Issues around forgetting that it was on the charger for a week at a time, physical limitations in using the Fitbit and not being able to wear it due to hospitalizations were all present in this study (see Section 5.3). Without the data triangulation, we would not have gained insights into the challenges around adherence to care plans and research study activities. For HCI and design studies wanting to explore CLH, designing for populations with cognitive impairments is essential. This might require adapting methods to support memory or cognitive impairments or designing protocols in a way that is flexible from a time perspective to adapt to aspects like hospitalizations or serious health events, allowing participants to pause their involvement until acute health issues resolve. Thus, future health technologies need modalities that allow for greater deviation and personalization to support challenges (anticipated and unforeseen) that may arise.

When participants are non-adherent, it is often seen in a binary light--yes or no. However, taking a human-centered data science approach to research design and data analysis, we shed light on how this can lead to false assumptions and completely incorrect assessments \cite{Guha}. The modern healthcare system is a complex sociotechnical ecosystem, and thus vulnerable to biases, assumptions, and data integrity issues. Through the interviews, we validated a myriad of reasons why people did not take their COVID vaccine as requested. It was not as simple as non-belief in science or religious and/or political reservations. Additionally, the complexity around compliance with wearing the Fitbit was highly nuanced, from individual challenges with functional use of technology, to not finding an ideal way to integrate the device use into everyday life, to being frustrated with technological issues (see Section 5.3, 5.4.1, 5.4.3) . If we are only assessing one aspect of this complex data ecosystem and trying to derive knowledge from it, we are potentially inducing more bias into our understanding.

Prior to the design of this research study, some research team members shadowed the clinicians at the PPCC, allowing for a deeper understanding of aspects of the clinic that would have been invisible otherwise. Because of this contextually rich, in-depth understanding of the environment and the patients \cite{Geertz}, we had unique insights that were used during the design of the research study. We were also very clear on what data the clinic was already collecting (e.g., patient-reported symptom severity) and the structured clinical notes that were present for all patients which informed the design of the data collection tools for this study. Thus, being deeply embedded and working closely with both patients and clinicians in the PPCC helped us build a deep understanding of individual needs as well as more complex challenges such as care plan adherence. 

\subsection{Implications for Triangulating Health Data}
Within the health informatics field, it is common to have multiple sources of health data integrated into the study design \cite{Weiskopf,Carney}. Previous HCI-focused health research has integrated EHR data, surveys, and patient-reported outcomes to better understand various phenomena like eating disorders \cite{Pater2019}, but also identified pros (e.g., deeper and more grounded understanding) and cons (e.g., complexity and resources needed to procure data) of doing so. In this work, the research team was embedded in the healthcare system, giving them a unique advantage to connect objective health data with subjective patient-reported outcomes. Therefore, assumptions within the field were examined in a manner that address issues like limitations associated with using social media analysis in healthcare \cite{Moorhead} and data quality of EHRs \cite{Kohane}. 

For this research, we had four separate data streams: EHR, Fitbit, surveys, and interviews. We posit this gives us a more holistic and realistic view of the individual patient--something that the EHR alone is not be able to provide \cite{Kohane}. Just looking at the EHR or the Fitbit data, both considered objective from a clinical standpoint, often will not give enough context about a patient's life experiences and how this has shaped their health behaviors and adherence to clinical recommendations and treatments. For example, we were able to identify adherence with clinical recommendations through the analysis of clinical encounters found in the EHR which, in some cases, was found to be contradictory to survey response data. Likewise, the interviews provided additional insights into non-adherence, illuminating aspects related to insurance/cost issues and the decision to focus on another family member's health before their own. Since we have the privilege of having access to EHR data, it allows us to directly assess ongoing healthcare utilization from an objective perspective. Future work is needed to isolate certain lifestyle and/or environmental factors to better understand their impacts on CLH-related symptoms, further expanding the levels of data triangulation. While the sleep and activity data didn't have direct correlations to better overall health in this study, the small sample size and lack of a baseline to compare it to might challenge any assumption that might be made from the  data presented.

\subsection{Operationalizing Research Findings}
Even though we are just starting to learn about CLH, consumer-facing technologies to support this condition have started to emerge. One example is the Stronger Together app, developed specifically to "support COVID-19 patients and long-haulers with their unique health needs" \footnote{https://www.cbc.ca/news/canada/hamilton/covid19-long-haulers-app-1.5904864}. This app addresses a core issue for CLH patients -- peer support. As discussed above, peer support is currently being found through online media \cite{Pater23,Day}, however there is no formal vetting of resources being shared or the levels of support that are actually seen in these spaces.  Thus, our research provides timely insights into needs and contexts of the targeted end-user as the marketplace rushes to create digital spaces and technology to support individuals living with CLH. 

These findings also have insights for the clinical field. A known pain point in U.S. healthcare is the ever increasing amounts of data being generated within the EHR \cite{Zulman}. Although technically feasible, this data deluge cannot be resolved by developing additional dashboards for clinicians who are already overwhelmed with information \cite{Jalilian}. Our research, which is embedded within a healthcare system and conducted in collaboration with clinicians, is sensitive to this burden of data overload. One key aspect driving this deluge of data is the need for shared decision making, especially with complex chronic illnesses that are multi-faceted \cite{Kooij}. Shared decision making is a key benefit of the EHR. However, in practice, issues like data retrieval, lack of visibility, and incomplete data erode the ability to maximize the impact of this benefit \cite{Alami}. Moreover, it is important to examine how providers interpret \& use information from different data streams and what implications it has for visualizations facilitating effective patient-provider communication around CLH. Thus, future research is needed to explore various types of data visualization to ensure that it is presenting what patients want their providers to know about their dynamic and evolving health status, yet done in a way that is clinically valuable to the providers and can be shown to have positive impacts on health outcomes.

Simply creating additional dashboards with objective patient data is not enough \cite{Ancker}. Through the triangulation of subjective patient-reported data and objective data from the EHR, we provide a contextually rich understanding of patients' overall health. Through this enhanced communication, we present a more complete picture of health indicators like mental health and social determinants of health and their impacts on patients. Commercial technologies can help collect and parse subjective data; however, providers often question the validity of this data as it is not being collected in a controlled environment \cite{Brown}. Thus, more education is needed on the clinical side as to the integrity and validity of data collected via commercial off-the-shelf technologies like fitness trackers, smart scales, and on-body trackers. Furthermore, EHR ecosystems need more conduits for the secure integration of non-clinical data and would need to consider data standards and appropriate policies \cite{Espinoza,Lobach,Koren}.

\subsection{Limitations \& Ethical Considerations}
The patient panel for this particular clinic skews heavily towards White/Caucasian, middle-aged female patients which is reflected in the patients we were able to recruit for this study. At the time of this writing, the PPCC patient panel of 884 patients is 90.0\% White/Caucasian, 72.3\%  female (average age = 51.5 years). Our study participants were also majority female, slightly younger (average age 49.1), and in-line with the racial makeup of the larger PPCC patient panel. Moreover, the physical location of the Parkview Health system is a potential factor contributing to the racial demographics of our study participants.

During the interviews, we asked participants to reflect on the impacts of CLH on their daily lives, potentially making them uncomfortable. A research nurse was present for each interview and when participants would mention concerning emotions or medical issues, she would engage them about these aspects. During several of the interviews this became valuable as participants had clinically-focused questions and became emotional when discussing mental health issues, for all of which the research nurse was able to provide immediate guidance and support. We maintained privacy by removing any personally identifiable information and quoting only relevant excerpts from participant interviews.

We also acknowledge the possibility of missing data in the system (e.g., patients not reporting impacts on mental health in their top 5 symptoms, incomplete or non-timely surveys) which can impact our overall understanding of lived experiences of CLH. Additionally, there is no activity/sleep data to compare patients' pre-COVID patterns. For this analysis, we assume that all EHR data is complete/whole. However, there is the potential that there is missing data due to information being documented in the wrong place or not-yet being entered into the EHR, which is a common limitation. For the healthcare utilization assessment, we only had access to encounters within the Parkview Health system unless a notation was made in the EHR about patients seeking care outside of the system, which we saw with one participant having healthcare appointments at the local Veterans Affairs offices. 

Finally, due to the embedded nature of the research team, specific patient sensitivities and data collection were made possible. Understanding both explicit and implicit barriers related to patients and data within the EHR allowed us to design a study where data collected via survey did not duplicate what was being collected clinically. Our access to the EHR also allowed us to directly connect digital traces (as expressed through Fitbit data), patient reported outcomes (surveys, interviews) and EHR data in a way that allows for a more holistic understanding of this chronic health issue and lived experiences associated with it.

 \section{Conclusion}
The rapid onset of COVID-19 stressed the infrastructure of healthcare, thus creating an urgent need for research that comprehensively characterizes the ongoing ramifications of the pandemic, including CLH. CLH is an emerging illness that continues to evolve in its clinical presentation and ongoing impact on the personal and professional lives of individuals. Current approaches to defining CLH rely mainly on survey or EHR data, presenting a singular dimension of the illness. In this study, we provide multi-dimensional data from a three-month study that chronicles the evolving nature of CLH symptomology within a cohort of patients. When thinking about how we might design systems that support patients' CLH journeys, including interactions with the health system, we must think about the limitations of current sociotechnical systems while, at the same time, design agile systems that take into consideration the variability of lived experiences of CLH. 

\begin{acks}
   We would like to acknowledge the anonymous monetary gift made to the Parkview Foundation that was given to directly support research within the PPCC. Additional support was provided through NSERC Discovery Early Career Researcher Grant RGPIN-2022-04570. 
\end{acks}

\bibliographystyle{ACM-Reference-Format}
\bibliography{references.bib}


\begin{thebibliography}{124}


\ifx \showCODEN    \undefined \def \showCODEN     #1{\unskip}     \fi
\ifx \showDOI      \undefined \def \showDOI       #1{#1}\fi
\ifx \showISBNx    \undefined \def \showISBNx     #1{\unskip}     \fi
\ifx \showISBNxiii \undefined \def \showISBNxiii  #1{\unskip}     \fi
\ifx \showISSN     \undefined \def \showISSN      #1{\unskip}     \fi
\ifx \showLCCN     \undefined \def \showLCCN      #1{\unskip}     \fi
\ifx \shownote     \undefined \def \shownote      #1{#1}          \fi
\ifx \showarticletitle \undefined \def \showarticletitle #1{#1}   \fi
\ifx \showURL      \undefined \def \showURL       {\relax}        \fi
\providecommand\bibfield[2]{#2}
\providecommand\bibinfo[2]{#2}
\providecommand\natexlab[1]{#1}
\providecommand\showeprint[2][]{arXiv:#2}

\bibitem[nat(2022)]%
        {national_heart}
 \bibinfo{year}{2022}\natexlab{}.
\newblock \bibinfo{title}{How Much Sleep is Enough?}
\newblock
\newblock
\urldef\tempurl%
\url{https://www.nhlbi.nih.gov/health/sleep/how-much-sleep#:~:text=Experts%20recommend%20that%20adults%20sleep,or%20more%20hours%20a%20night.}
\showURL{%
\tempurl}


\bibitem[ad~Kristen~Miller(2011)]%
        {Gordon}
\bibfield{author}{\bibinfo{person}{Gordon B.~Willis ad Kristen~Miller}.} \bibinfo{year}{2011}\natexlab{}.
\newblock \showarticletitle{Cross-cultural cognitive interviewing: Seeking comparability and enhancing understanding}.
\newblock \bibinfo{journal}{\emph{Field Methods}} \bibinfo{volume}{23}, \bibinfo{number}{4} (\bibinfo{year}{2011}), \bibinfo{pages}{331--341}.
\newblock
\urldef\tempurl%
\url{https://doi.org/10.1177/1525822X11416092}
\showDOI{\tempurl}


\bibitem[Alami et~al\mbox{.}(2020)]%
        {Alami}
\bibfield{author}{\bibinfo{person}{Jawad Alami}, \bibinfo{person}{Stephen Borowitz}, {and} \bibinfo{person}{Sara~L. Riggs}.} \bibinfo{year}{2020}\natexlab{}.
\newblock \showarticletitle{Usability Challenges with EHRs During Pre-Rounding in the Pediatric Acute Care Department}.
\newblock \bibinfo{journal}{\emph{Proceedings of the Human Factors and Ergonomics Society Annual Meeting}} \bibinfo{volume}{64}, \bibinfo{number}{1} (\bibinfo{year}{2020}), \bibinfo{pages}{1282--1286}.
\newblock
\urldef\tempurl%
\url{https://doi.org/10.1177/1071181320641305}
\showDOI{\tempurl}


\bibitem[Ancker et~al\mbox{.}(2017)]%
        {Ancker}
\bibfield{author}{\bibinfo{person}{Jessica~S. Ancker}, \bibinfo{person}{Alison Edwards}, \bibinfo{person}{Sarah Nosal}, \bibinfo{person}{Diane Hauser}, \bibinfo{person}{Elizabeth Mauer}, \bibinfo{person}{Rainu Kaushal}, {and} \bibinfo{person}{HITEC Investigators}.} \bibinfo{year}{2017}\natexlab{}.
\newblock \showarticletitle{Effects of workload, work complexity, and repeated alerts on alert fatigue in a clinical decision support system}.
\newblock \bibinfo{journal}{\emph{BMC medical informatics and decision making}}  \bibinfo{volume}{17} (\bibinfo{year}{2017}), \bibinfo{pages}{1--9}.
\newblock
\urldef\tempurl%
\url{https://doi.org/10.1186/s12911-017-0430-8}
\showDOI{\tempurl}


\bibitem[Antony et~al\mbox{.}(2023)]%
        {Antony}
\bibfield{author}{\bibinfo{person}{Blessy Antony}, \bibinfo{person}{Hannah Blau}, \bibinfo{person}{Elena Casiraghi}, \bibinfo{person}{Johanna~J. Loomba}, \bibinfo{person}{Tiffany~J. Callahan}, \bibinfo{person}{Bryan~J. Laraway}, \bibinfo{person}{Kenneth~J. Wilkins}, \bibinfo{person}{Corneliu~C. Antonescu}, \bibinfo{person}{Giorgio Valenti}, \bibinfo{person}{Andrew~E. Williams}, \bibinfo{person}{Peter~N. Robinson}, \bibinfo{person}{Justin~T. Resse}, {and} \bibinfo{person}{T.M Murali}.} \bibinfo{year}{2023}\natexlab{}.
\newblock \showarticletitle{Predictive models of long COVID}.
\newblock \bibinfo{journal}{\emph{EBioMedicine}}  \bibinfo{volume}{96} (\bibinfo{year}{2023}), \bibinfo{pages}{104777}.
\newblock
\urldef\tempurl%
\url{https://doi.org/10.1016/j.ebiom.2023.104777}
\showDOI{\tempurl}


\bibitem[Aragon et~al\mbox{.}(2022)]%
        {Guha}
\bibfield{author}{\bibinfo{person}{Cecilia Aragon}, \bibinfo{person}{Shion Guha}, \bibinfo{person}{Marina Kogan}, \bibinfo{person}{Michael Miuller}, {and} \bibinfo{person}{Gina Neff}.} \bibinfo{year}{2022}\natexlab{}.
\newblock \bibinfo{booktitle}{\emph{Human-Centered Data Science: An Introduction}}.
\newblock \bibinfo{publisher}{MIT Press}, \bibinfo{address}{Boston}.
\newblock


\bibitem[Bardzell et~al\mbox{.}(2019)]%
        {Bardzell}
\bibfield{author}{\bibinfo{person}{Jeffrey Bardzell}, \bibinfo{person}{Shaowen Bardzell}, \bibinfo{person}{Amanda Lazar}, {and} \bibinfo{person}{Norman~Makoto Su}.} \bibinfo{year}{2019}\natexlab{}.
\newblock \showarticletitle{(Re-)Framing Menopause Experiences for HCI and Design}. In \bibinfo{booktitle}{\emph{Proceedings of the 2019 CHI Conference on Human Factors in Computing Systems}} (Glasgow, Scotland Uk) \emph{(\bibinfo{series}{CHI '19})}. \bibinfo{publisher}{Association for Computing Machinery}, \bibinfo{address}{New York, NY, USA}, \bibinfo{pages}{1–13}.
\newblock
\showISBNx{9781450359702}
\urldef\tempurl%
\url{https://doi.org/10.1145/3290605.3300345}
\showDOI{\tempurl}


\bibitem[Barricelli et~al\mbox{.}(2016)]%
        {Barricelli}
\bibfield{author}{\bibinfo{person}{Barbara~Rita Barricelli}, \bibinfo{person}{Stefano Valtolina}, {and} \bibinfo{person}{Jose Abdelnour-Nocera}.} \bibinfo{year}{2016}\natexlab{}.
\newblock \showarticletitle{Sociotechnical Design of MHealth Applications for Chronic Diseases}. In \bibinfo{booktitle}{\emph{Proceedings of the 18th International Conference on Human-Computer Interaction with Mobile Devices and Services Adjunct}} (Florence, Italy) \emph{(\bibinfo{series}{MobileHCI '16})}. \bibinfo{publisher}{Association for Computing Machinery}, \bibinfo{address}{New York, NY, USA}, \bibinfo{pages}{1097–1100}.
\newblock
\showISBNx{9781450344135}
\urldef\tempurl%
\url{https://doi.org/10.1145/2957265.2965009}
\showDOI{\tempurl}


\bibitem[Bers et~al\mbox{.}(1998)]%
        {Bers}
\bibfield{author}{\bibinfo{person}{Marina~Umaschi Bers}, \bibinfo{person}{Edith Ackermann}, \bibinfo{person}{Justine Cassell}, \bibinfo{person}{Beth Donegan}, \bibinfo{person}{Joseph Gonzalez-Heydrich}, \bibinfo{person}{David~Ray DeMaso}, \bibinfo{person}{Carol Strohecker}, \bibinfo{person}{Sarah Lualdi}, \bibinfo{person}{Dennis Bromley}, {and} \bibinfo{person}{Judith Karlin}.} \bibinfo{year}{1998}\natexlab{}.
\newblock \showarticletitle{Interactive Storytelling Environments: Coping with Cardiac Illness at Boston's Children's Hospital}. In \bibinfo{booktitle}{\emph{Proceedings of the SIGCHI Conference on Human Factors in Computing Systems}} (Los Angeles, California, USA) \emph{(\bibinfo{series}{CHI '98})}. \bibinfo{publisher}{ACM Press/Addison-Wesley Publishing Co.}, \bibinfo{address}{USA}, \bibinfo{pages}{603–610}.
\newblock
\showISBNx{0201309874}
\urldef\tempurl%
\url{https://doi.org/10.1145/274644.274725}
\showDOI{\tempurl}


\bibitem[Birnholtz and Jones-Rounds(2010)]%
        {Birnholtz}
\bibfield{author}{\bibinfo{person}{Jeremy Birnholtz} {and} \bibinfo{person}{McKenzie Jones-Rounds}.} \bibinfo{year}{2010}\natexlab{}.
\newblock \showarticletitle{Independence and Interaction: Understanding Seniors' Privacy and Awareness Needs for Aging in Place}. In \bibinfo{booktitle}{\emph{Proceedings of the SIGCHI Conference on Human Factors in Computing Systems}} (Atlanta, Georgia, USA) \emph{(\bibinfo{series}{CHI '10})}. \bibinfo{publisher}{Association for Computing Machinery}, \bibinfo{address}{New York, NY, USA}, \bibinfo{pages}{143–152}.
\newblock
\showISBNx{9781605589299}
\urldef\tempurl%
\url{https://doi.org/10.1145/1753326.1753349}
\showDOI{\tempurl}


\bibitem[Bohn et~al\mbox{.}(2022)]%
        {Bohn}
\bibfield{author}{\bibinfo{person}{Camden Bohn}, \bibinfo{person}{Jason Li}, \bibinfo{person}{Noah Todd}, \bibinfo{person}{Jessica Pater}, \bibinfo{person}{Jeanne Carroll}, \bibinfo{person}{Brian Henriksen}, {and} \bibinfo{person}{Fen-Lei Chang}.} \bibinfo{year}{2022}\natexlab{}.
\newblock \showarticletitle{Predicting Cognitive Impairment in Long-COVID Patients: A Demographic and Comorbid Analysis using BrainCheck Cognitive Assessment}.
\newblock \bibinfo{journal}{\emph{Proceedings of IMPRS}} \bibinfo{volume}{5}, \bibinfo{number}{1} (\bibinfo{year}{2022}).
\newblock


\bibitem[Bossen et~al\mbox{.}(2012)]%
        {Bossen12}
\bibfield{author}{\bibinfo{person}{Claus Bossen}, \bibinfo{person}{Lotte Groth~Jensen}, {and} \bibinfo{person}{Flemming Witt}.} \bibinfo{year}{2012}\natexlab{}.
\newblock \showarticletitle{Medical Secretaries' Care of Records: The Cooperative Work of a Non-Clinical Group}. In \bibinfo{booktitle}{\emph{Proceedings of the ACM 2012 Conference on Computer Supported Cooperative Work}} \emph{(\bibinfo{series}{CSCW '12})}. \bibinfo{publisher}{Association for Computing Machinery}, \bibinfo{address}{New York, NY, USA}, \bibinfo{pages}{921–930}.
\newblock
\showISBNx{9781450310864}
\urldef\tempurl%
\url{https://doi.org/10.1145/2145204.2145341}
\showDOI{\tempurl}


\bibitem[Bossen and Jensen(2014)]%
        {Bossen}
\bibfield{author}{\bibinfo{person}{Claus Bossen} {and} \bibinfo{person}{Lotte~Groth Jensen}.} \bibinfo{year}{2014}\natexlab{}.
\newblock \showarticletitle{How Physicians 'Achieve Overview': A Case-Based Study in a Hospital Ward}. In \bibinfo{booktitle}{\emph{Proceedings of the 17th ACM Conference on Computer Supported Cooperative Work \& Social Computing}} (Baltimore, Maryland, USA) \emph{(\bibinfo{series}{CSCW '14})}. \bibinfo{publisher}{Association for Computing Machinery}, \bibinfo{address}{New York, NY, USA}, \bibinfo{pages}{257–268}.
\newblock
\showISBNx{9781450325400}
\urldef\tempurl%
\url{https://doi.org/10.1145/2531602.2531620}
\showDOI{\tempurl}


\bibitem[Bowman(2013)]%
        {Bowman13}
\bibfield{author}{\bibinfo{person}{Sue Bowman}.} \bibinfo{year}{2013}\natexlab{}.
\newblock \showarticletitle{Impact of electronic health record systems on information integrity: quality and safety implications}.
\newblock \bibinfo{journal}{\emph{Perspectives in health information management}} \bibinfo{volume}{10}, \bibinfo{number}{Fall} (\bibinfo{year}{2013}), \bibinfo{pages}{1c}.
\newblock


\bibitem[Braun and Clarke(2019)]%
        {braun2019reflecting}
\bibfield{author}{\bibinfo{person}{Virginia Braun} {and} \bibinfo{person}{Victoria Clarke}.} \bibinfo{year}{2019}\natexlab{}.
\newblock \showarticletitle{Reflecting on reflexive thematic analysis}.
\newblock \bibinfo{journal}{\emph{Qualitative research in sport, exercise and health}} \bibinfo{volume}{11}, \bibinfo{number}{4} (\bibinfo{year}{2019}), \bibinfo{pages}{589--597}.
\newblock


\bibitem[Brown(2019)]%
        {Brown}
\bibfield{author}{\bibinfo{person}{Dalvin Brown}.} \bibinfo{year}{2019}\natexlab{}.
\newblock \bibinfo{title}{Doctors say most metrics provided by your Apple Watch, Fitbit aren't helpful to them}.
\newblock
\newblock
\urldef\tempurl%
\url{https://www.usatoday.com/story/tech/2019/08/14/how-doctors-really-feel-data-your-apple-watch-fitbit/1900968001/}
\showURL{%
\tempurl}


\bibitem[Cadmus-Bertram et~al\mbox{.}(2023)]%
        {Cadmus}
\bibfield{author}{\bibinfo{person}{L. Cadmus-Bertram}, \bibinfo{person}{P. Solk}, \bibinfo{person}{M. Agnew}, \bibinfo{person}{J. Starikovsky}, \bibinfo{person}{C. Schmidt}, \bibinfo{person}{W.A. Morelli}, {and} \bibinfo{person}{S.M. Phillips}.} \bibinfo{year}{2023}\natexlab{}.
\newblock \showarticletitle{A multi-site trial of an electronic health integrated physical activity promotion intervention in breast and endometrial cancers survivors: MyActivity study protocol}.
\newblock \bibinfo{journal}{\emph{Contemporary Clinical Trials}}  \bibinfo{volume}{130} (\bibinfo{year}{2023}), \bibinfo{pages}{107187}.
\newblock


\bibitem[Cajander and Gr\"{u}nloh(2019)]%
        {Cajander19}
\bibfield{author}{\bibinfo{person}{\r{A}sa Cajander} {and} \bibinfo{person}{Christiane Gr\"{u}nloh}.} \bibinfo{year}{2019}\natexlab{}.
\newblock \showarticletitle{Electronic Health Records Are More Than a Work Tool: Conflicting Needs of Direct and Indirect Stakeholders}. In \bibinfo{booktitle}{\emph{Proceedings of the 2019 CHI Conference on Human Factors in Computing Systems}} (Glasgow, Scotland Uk) \emph{(\bibinfo{series}{CHI '19})}. \bibinfo{publisher}{Association for Computing Machinery}, \bibinfo{address}{New York, NY, USA}, \bibinfo{pages}{1–13}.
\newblock
\showISBNx{9781450359702}
\urldef\tempurl%
\url{https://doi.org/10.1145/3290605.3300865}
\showDOI{\tempurl}


\bibitem[Carbajal and Gleeson(2022)]%
        {COVID}
\bibfield{author}{\bibinfo{person}{Erica Carbajal} {and} \bibinfo{person}{Cailey Gleeson}.} \bibinfo{year}{2022}\natexlab{}.
\newblock \bibinfo{title}{66 hospitals, health systems that have launched post-COVID-19 clinics}.
\newblock \bibinfo{howpublished}{https://www.beckershospitalreview.com/patient-safety-outcomes/13-hospitals-health-systems-that-have-launched-post-covid-19-clinics.html}.
\newblock


\bibitem[Carney and Kong(2017)]%
        {Carney}
\bibfield{author}{\bibinfo{person}{Timothy~Jay Carney} {and} \bibinfo{person}{Amanda~Y. Kong}.} \bibinfo{year}{2017}\natexlab{}.
\newblock \showarticletitle{Leveraging health informatics to foster a smart systems response to health disparities and health equity challenges}.
\newblock \bibinfo{journal}{\emph{Journal of biomedical informatics}}  \bibinfo{volume}{68} (\bibinfo{year}{2017}), \bibinfo{pages}{184--188}.
\newblock
\urldef\tempurl%
\url{https://doi.org/10.1016/j.jbi.2017.02.011}
\showDOI{\tempurl}


\bibitem[Carter et~al\mbox{.}(2014)]%
        {Carter}
\bibfield{author}{\bibinfo{person}{N. Carter}, \bibinfo{person}{D. Bryant-Lukosius}, \bibinfo{person}{A. DiCenso}, \bibinfo{person}{J. Blythe}, {and} \bibinfo{person}{A.J. Neville}.} \bibinfo{year}{2014}\natexlab{}.
\newblock \showarticletitle{The use of triangulation in qualitative research}.
\newblock \bibinfo{journal}{\emph{Oncol Nurs Forum}} \bibinfo{volume}{41}, \bibinfo{number}{5} (\bibinfo{year}{2014}), \bibinfo{pages}{545--547}.
\newblock
\urldef\tempurl%
\url{https://doi.org/10.1188/14.ONF.545-547}
\showDOI{\tempurl}


\bibitem[Cha et~al\mbox{.}(2022)]%
        {Cha}
\bibfield{author}{\bibinfo{person}{Yoon~Jeong Cha}, \bibinfo{person}{Arpita Saxena}, \bibinfo{person}{Alice Wou}, \bibinfo{person}{Joyce Lee}, \bibinfo{person}{Mark~W Newman}, {and} \bibinfo{person}{Sun~Young Park}.} \bibinfo{year}{2022}\natexlab{}.
\newblock \showarticletitle{Transitioning Toward Independence: Enhancing Collaborative Self-Management of Children with Type 1 Diabetes}. In \bibinfo{booktitle}{\emph{Proceedings of the 2022 CHI Conference on Human Factors in Computing Systems}} (New Orleans, LA, USA) \emph{(\bibinfo{series}{CHI '22})}. \bibinfo{publisher}{Association for Computing Machinery}, \bibinfo{address}{New York, NY, USA}, Article \bibinfo{articleno}{522}, \bibinfo{numpages}{17}~pages.
\newblock
\showISBNx{9781450391573}
\urldef\tempurl%
\url{https://doi.org/10.1145/3491102.3502055}
\showDOI{\tempurl}


\bibitem[Chaudhry et~al\mbox{.}(2016)]%
        {Chaudhry}
\bibfield{author}{\bibinfo{person}{Beenish~M. Chaudhry}, \bibinfo{person}{Christopher Schaefbauer}, \bibinfo{person}{Ben Jelen}, \bibinfo{person}{Katie~A. Siek}, {and} \bibinfo{person}{Kay Connelly}.} \bibinfo{year}{2016}\natexlab{}.
\newblock \showarticletitle{Evaluation of a Food Portion Size Estimation Interface for a Varying Literacy Population}. In \bibinfo{booktitle}{\emph{Proceedings of the 2016 CHI Conference on Human Factors in Computing Systems}} (San Jose, California, USA) \emph{(\bibinfo{series}{CHI '16})}. \bibinfo{publisher}{Association for Computing Machinery}, \bibinfo{address}{New York, NY, USA}, \bibinfo{pages}{5645–5657}.
\newblock
\urldef\tempurl%
\url{https://doi.org/10.1145/2858036.2858554}
\showDOI{\tempurl}


\bibitem[Chen et~al\mbox{.}(2022)]%
        {Chen}
\bibfield{author}{\bibinfo{person}{Chen Chen}, \bibinfo{person}{Spencer~R. Haupert}, \bibinfo{person}{Lauren Zimmermann}, \bibinfo{person}{Xu Sho}, \bibinfo{person}{Lars~G. Fritsche}, {and} \bibinfo{person}{Bhramar Mukherjee}.} \bibinfo{year}{2022}\natexlab{}.
\newblock \showarticletitle{Global Prevalence of Post COVID-19 Condition or Long COVID: A Meta-Analysis and Systematic Review}.
\newblock \bibinfo{journal}{\emph{The Journal of Infectious Diseases}}  \bibinfo{volume}{136} (\bibinfo{year}{2022}), \bibinfo{pages}{1593--1607}.
\newblock
\urldef\tempurl%
\url{https://doi.org/10.1093/infdis/jiac136}
\showDOI{\tempurl}


\bibitem[Cherenshchykova and Miller(2021)]%
        {Miller}
\bibfield{author}{\bibinfo{person}{Anna Cherenshchykova} {and} \bibinfo{person}{Andrew~D. Miller}.} \bibinfo{year}{2021}\natexlab{}.
\newblock \showarticletitle{Sociotechnical Design Opportunities for Pervasive Family Sleep Technologies}. In \bibinfo{booktitle}{\emph{Proceedings of the 14th EAI International Conference on Pervasive Computing Technologies for Healthcare}} \emph{(\bibinfo{series}{PervasiveHealth '20})}. \bibinfo{publisher}{Association for Computing Machinery}, \bibinfo{address}{New York, NY, USA}, \bibinfo{pages}{11--20}.
\newblock
\urldef\tempurl%
\url{https://doi.org/10.1145/3290605.3300865}
\showDOI{\tempurl}


\bibitem[Clynch and Kellett(2015)]%
        {Clynch15}
\bibfield{author}{\bibinfo{person}{Neil Clynch} {and} \bibinfo{person}{John Kellett}.} \bibinfo{year}{2015}\natexlab{}.
\newblock \showarticletitle{Medical documentation: part of the solution, or part of the problem? A narrative review of the literature on the time spent on and value of medical documentation}.
\newblock \bibinfo{journal}{\emph{International journal of medical informatics}} \bibinfo{volume}{84}, \bibinfo{number}{4} (\bibinfo{year}{2015}), \bibinfo{pages}{221--228}.
\newblock
\urldef\tempurl%
\url{https://doi.org/10.1016/j.ijmedinf.2014.12.001}
\showDOI{\tempurl}


\bibitem[Cohen and Manion(254)]%
        {Cohen}
\bibfield{author}{\bibinfo{person}{L. Cohen} {and} \bibinfo{person}{L. Manion}.} \bibinfo{year}{254}\natexlab{}.
\newblock \bibinfo{booktitle}{\emph{Research methods in education} (\bibinfo{edition}{5th} ed.)}.
\newblock \bibinfo{publisher}{Routledge}, \bibinfo{address}{London}.
\newblock


\bibitem[Collins(2021)]%
        {Collins}
\bibfield{author}{\bibinfo{person}{Frances Collins}.} \bibinfo{year}{2021}\natexlab{}.
\newblock \bibinfo{title}{NIH launches new initiative to study “Long COVID}.
\newblock
\newblock
\urldef\tempurl%
\url{https://www.nih.gov/about-nih/who-we-are/nih-director/statements/nih-launches-new-initiative-study-long-covid}
\showURL{%
\tempurl}


\bibitem[Corbin and Strauss(2015)]%
        {corbin2014basics}
\bibfield{author}{\bibinfo{person}{Juliet Corbin} {and} \bibinfo{person}{Anselm Strauss}.} \bibinfo{year}{2015}\natexlab{}.
\newblock \bibinfo{booktitle}{\emph{Basics of qualitative research: Techniques and procedures for developing grounded theory} (\bibinfo{edition}{14} ed.)}.
\newblock \bibinfo{publisher}{Sage publications}, \bibinfo{address}{Thousand Oaks, CA}.
\newblock


\bibitem[Corman et~al\mbox{.}(2022)]%
        {Corman}
\bibfield{author}{\bibinfo{person}{B.~H.~P. Corman}, \bibinfo{person}{S. Rajupet}, \bibinfo{person}{F. Ye}, {and} \bibinfo{person}{E.R. Schoenfield}.} \bibinfo{year}{2022}\natexlab{}.
\newblock \showarticletitle{The Role of Unobtrusive Home-Based Continuous Sensing in the Management of Postacute Sequelae of SARS CoV-2}.
\newblock \bibinfo{journal}{\emph{Journal of Medical Internet Research}} \bibinfo{volume}{24}, \bibinfo{number}{1} (\bibinfo{year}{2022}), \bibinfo{pages}{e32713}.
\newblock
\urldef\tempurl%
\url{https://doi.org/10.2196/32713}
\showDOI{\tempurl}


\bibitem[Cubranic et~al\mbox{.}(1998)]%
        {Cubranic}
\bibfield{author}{\bibinfo{person}{Davor Cubranic}, \bibinfo{person}{Kellogg~S. Booth}, {and} \bibinfo{person}{Kate Collie}.} \bibinfo{year}{1998}\natexlab{}.
\newblock \showarticletitle{Computer Support for Distance Art Therapy}. In \bibinfo{booktitle}{\emph{CHI 98 Conference Summary on Human Factors in Computing Systems}} (Los Angeles, California, USA) \emph{(\bibinfo{series}{CHI '98})}. \bibinfo{publisher}{Association for Computing Machinery}, \bibinfo{address}{New York, NY, USA}, \bibinfo{pages}{277–278}.
\newblock
\showISBNx{1581130287}
\urldef\tempurl%
\url{https://doi.org/10.1145/286498.286758}
\showDOI{\tempurl}


\bibitem[Dalko et~al\mbox{.}(2023)]%
        {Dalko}
\bibfield{author}{\bibinfo{person}{Katharina Dalko}, \bibinfo{person}{Bernhard Kraft}, \bibinfo{person}{Patrick Jahn}, \bibinfo{person}{Jan Schildmann}, {and} \bibinfo{person}{Sebastian Hofstetter}.} \bibinfo{year}{2023}\natexlab{}.
\newblock \showarticletitle{Cocreation of Assistive Technologies for Patients With Long COVID: Qualitative Analysis of a Literature Review on the Challenges of Patient Involvement in Health and Nursing Sciences}.
\newblock \bibinfo{journal}{\emph{Journal of Medical Internet Research}}  \bibinfo{volume}{25} (\bibinfo{year}{2023}), \bibinfo{pages}{e46297}.
\newblock
\urldef\tempurl%
\url{https://doi.org/10.2196/46297}
\showDOI{\tempurl}


\bibitem[Das~Swain et~al\mbox{.}(2022)]%
        {Vedant}
\bibfield{author}{\bibinfo{person}{Vedant Das~Swain}, \bibinfo{person}{Victor Chen}, \bibinfo{person}{Shrija Mishra}, \bibinfo{person}{Stephen~M. Mattingly}, \bibinfo{person}{Gregory~D. Abowd}, {and} \bibinfo{person}{Munmun De~Choudhury}.} \bibinfo{year}{2022}\natexlab{}.
\newblock \showarticletitle{Semantic Gap in Predicting Mental Wellbeing through Passive Sensing}. In \bibinfo{booktitle}{\emph{Proceedings of the 2022 CHI Conference on Human Factors in Computing Systems}} (, New Orleans, LA, USA,) \emph{(\bibinfo{series}{CHI '22})}. \bibinfo{publisher}{Association for Computing Machinery}, \bibinfo{address}{New York, NY, USA}, Article \bibinfo{articleno}{374}, \bibinfo{numpages}{16}~pages.
\newblock
\showISBNx{9781450391573}
\urldef\tempurl%
\url{https://doi.org/10.1145/3491102.3502037}
\showDOI{\tempurl}


\bibitem[Day(2022)]%
        {Day}
\bibfield{author}{\bibinfo{person}{HLS Day}.} \bibinfo{year}{2022}\natexlab{}.
\newblock \showarticletitle{Exploring Online Peer Support Groups for Adults Experiencing Long COVID in the United Kingdom: Qualitative Interview Study}.
\newblock \bibinfo{journal}{\emph{Journal of Medical Internet Research}} \bibinfo{volume}{24}, \bibinfo{number}{5} (\bibinfo{year}{2022}), \bibinfo{pages}{e37674}.
\newblock
\urldef\tempurl%
\url{https://doi.org/10.2196/37674}
\showDOI{\tempurl}


\bibitem[De~Croon et~al\mbox{.}(2021)]%
        {Verbert}
\bibfield{author}{\bibinfo{person}{Robin~De De~Croon}, \bibinfo{person}{Artuur Leeuwenberg}, \bibinfo{person}{Jan Aerts}, \bibinfo{person}{Marie-Francine Moens}, \bibinfo{person}{Vero~Vanden Vanden~Abeele}, {and} \bibinfo{person}{Katrien Verbert}.} \bibinfo{year}{2021}\natexlab{}.
\newblock \showarticletitle{TIEVis: A Visual Analytics Dashboard for Temporal Information Extracted from Clinical Reports}. In \bibinfo{booktitle}{\emph{26th International Conference on Intelligent User Interfaces - Companion}} (College Station, TX, USA) \emph{(\bibinfo{series}{IUI '21 Companion})}. \bibinfo{publisher}{Association for Computing Machinery}, \bibinfo{address}{New York, NY, USA}, \bibinfo{pages}{34–36}.
\newblock
\showISBNx{9781450380188}
\urldef\tempurl%
\url{https://doi.org/10.1145/3397482.3450731}
\showDOI{\tempurl}


\bibitem[Ding et~al\mbox{.}(2021)]%
        {Ding}
\bibfield{author}{\bibinfo{person}{Xianghua~(Sharon) Ding}, \bibinfo{person}{Shuhan Wei}, \bibinfo{person}{Xinning Gui}, \bibinfo{person}{Ning Gu}, {and} \bibinfo{person}{Peng Zhang}.} \bibinfo{year}{2021}\natexlab{}.
\newblock \showarticletitle{Data Engagement Reconsidered: A Study of Automatic Stress Tracking Technology in Use}. In \bibinfo{booktitle}{\emph{Proceedings of the 2021 CHI Conference on Human Factors in Computing Systems}} (Yokohama, Japan) \emph{(\bibinfo{series}{CHI '21})}. \bibinfo{publisher}{Association for Computing Machinery}, \bibinfo{address}{New York, NY, USA}, Article \bibinfo{articleno}{535}, \bibinfo{numpages}{13}~pages.
\newblock
\showISBNx{9781450380966}
\urldef\tempurl%
\url{https://doi.org/10.1145/3411764.3445763}
\showDOI{\tempurl}


\bibitem[Dinh-Le et~al\mbox{.}(2019)]%
        {Chokshi}
\bibfield{author}{\bibinfo{person}{Catherine Dinh-Le}, \bibinfo{person}{Rachel Chuang}, \bibinfo{person}{Sara Chokshi}, {and} \bibinfo{person}{Devin Mann}.} \bibinfo{year}{2019}\natexlab{}.
\newblock \showarticletitle{Wearable health technology and electronic health record integration: scoping review and future directions}.
\newblock \bibinfo{journal}{\emph{JMIR mHealth and uHealth}} \bibinfo{volume}{7}, \bibinfo{number}{9} (\bibinfo{year}{2019}), \bibinfo{pages}{e12861}.
\newblock
\urldef\tempurl%
\url{https://doi.org/10.2196/12861}
\showDOI{\tempurl}


\bibitem[Ehrenberg(2000)]%
        {Ehrenberg}
\bibfield{author}{\bibinfo{person}{Bruce Ehrenberg}.} \bibinfo{year}{2000}\natexlab{}.
\newblock \showarticletitle{Importance of sleep restoration in co-morbid disease: effect of anticonvulsants}.
\newblock \bibinfo{journal}{\emph{Neurology}} \bibinfo{volume}{54}, \bibinfo{number}{5} (\bibinfo{year}{2000}), \bibinfo{pages}{S33--S37}.
\newblock


\bibitem[Ernala et~al\mbox{.}(2019)]%
        {Ernala19}
\bibfield{author}{\bibinfo{person}{Sindhu~Kiranmai Ernala}, \bibinfo{person}{Michael~L. Birnbaum}, \bibinfo{person}{Kristin~A. Candan}, \bibinfo{person}{Asra~F. Rizvi}, \bibinfo{person}{William~A. Sterling}, \bibinfo{person}{John~M. Kane}, {and} \bibinfo{person}{Munmun De~Choudhury}.} \bibinfo{year}{2019}\natexlab{}.
\newblock \showarticletitle{Methodological Gaps in Predicting Mental Health States from Social Media: Triangulating Diagnostic Signals}. In \bibinfo{booktitle}{\emph{Proceedings of the 2019 CHI Conference on Human Factors in Computing Systems}} (Glasgow, Scotland Uk) \emph{(\bibinfo{series}{CHI '19})}. \bibinfo{publisher}{Association for Computing Machinery}, \bibinfo{address}{New York, NY, USA}, \bibinfo{pages}{1–16}.
\newblock
\showISBNx{9781450359702}
\urldef\tempurl%
\url{https://doi.org/10.1145/3290605.3300364}
\showDOI{\tempurl}


\bibitem[Ernala et~al\mbox{.}(2022)]%
        {Munmun}
\bibfield{author}{\bibinfo{person}{Sindhu~Kiranmai Ernala}, \bibinfo{person}{Jordyn Seybolt}, \bibinfo{person}{Dong~Whi Yoo}, \bibinfo{person}{Michael~L. Birnbaum}, \bibinfo{person}{John~M. Kane}, {and} \bibinfo{person}{Munmun De~Choudhury}.} \bibinfo{year}{2022}\natexlab{}.
\newblock \showarticletitle{The Reintegration Journey Following a Psychiatric Hospitalization: Examining the Role of Social Technologies}.
\newblock \bibinfo{journal}{\emph{Proc. ACM Hum.-Comput. Interact.}} \bibinfo{volume}{6}, \bibinfo{number}{CSCW1}, Article \bibinfo{articleno}{122} (\bibinfo{date}{apr} \bibinfo{year}{2022}), \bibinfo{numpages}{31}~pages.
\newblock
\urldef\tempurl%
\url{https://doi.org/10.1145/3512969}
\showDOI{\tempurl}


\bibitem[Espinoza et~al\mbox{.}(2023)]%
        {Espinoza}
\bibfield{author}{\bibinfo{person}{Juan Espinoza}, \bibinfo{person}{Nicole~Y. Xu}, \bibinfo{person}{Kevin~T. Nguyen}, {and} \bibinfo{person}{David~C. Klonoff}.} \bibinfo{year}{2023}\natexlab{}.
\newblock \showarticletitle{The need for data standards and implementation policies to integrate CGM data into the electronic health record}.
\newblock \bibinfo{journal}{\emph{Journal of Diabetes Science and Technology}} \bibinfo{volume}{17}, \bibinfo{number}{2} (\bibinfo{year}{2023}), \bibinfo{pages}{495--502}.
\newblock
\urldef\tempurl%
\url{https://doi.org/10.1177/19322968211058148}
\showDOI{\tempurl}


\bibitem[Fisher et~al\mbox{.}(2023)]%
        {Fisher}
\bibfield{author}{\bibinfo{person}{Luke Fisher}, \bibinfo{person}{Benecia Goka}, \bibinfo{person}{Jessica Pater}, {and} \bibinfo{person}{Fen-Lei Chang}.} \bibinfo{year}{2023}\natexlab{}.
\newblock \showarticletitle{Neutrophil-to-Lymphocyte Ratio (NLR) to Monitor Neuroinflammation Status During Long COVID}.
\newblock \bibinfo{journal}{\emph{Proceedings of IMPRS}} \bibinfo{volume}{6}, \bibinfo{number}{1} (\bibinfo{year}{2023}).
\newblock


\bibitem[for Disease~Control and Prevention(2023a)]%
        {CDC_end}
\bibfield{author}{\bibinfo{person}{Centers for Disease~Control} {and} \bibinfo{person}{Prevention}.} \bibinfo{year}{2023}\natexlab{a}.
\newblock \bibinfo{title}{End of the Federal COVID-19 Public Health Emergency (PHE) Declaration}.
\newblock
\newblock
\urldef\tempurl%
\url{https://www.cdc.gov/coronavirus/2019-ncov/your-health/end-of-phe.html}
\showURL{%
\tempurl}


\bibitem[for Disease~Control and Prevention(2023b)]%
        {CDC_CLH}
\bibfield{author}{\bibinfo{person}{Centers for Disease~Control} {and} \bibinfo{person}{Prevention}.} \bibinfo{year}{2023}\natexlab{b}.
\newblock \bibinfo{title}{Long COVID or Post-COVID Conditions}.
\newblock
\newblock
\urldef\tempurl%
\url{https://www.cdc.gov/coronavirus/2019-ncov/long-term-effects/index.html}
\showURL{%
\tempurl}


\bibitem[Fusch et~al\mbox{.}(2018)]%
        {Fusch}
\bibfield{author}{\bibinfo{person}{Patricia Fusch}, \bibinfo{person}{Gene~E. Fusch}, {and} \bibinfo{person}{Lawrence~R. Ness}.} \bibinfo{year}{2018}\natexlab{}.
\newblock \showarticletitle{Denzin’s paradigm shift: Revisiting triangulation in qualitative research}.
\newblock \bibinfo{journal}{\emph{Journal of Sustainable Social Change}} \bibinfo{volume}{10}, \bibinfo{number}{1} (\bibinfo{year}{2018}), \bibinfo{pages}{2}.
\newblock


\bibitem[Geetz(1973)]%
        {Geertz}
\bibfield{author}{\bibinfo{person}{Clifford Geetz}.} \bibinfo{year}{1973}\natexlab{}.
\newblock \bibinfo{booktitle}{\emph{Thick description: Toward an interpretive theory of culture}}.
\newblock \bibinfo{publisher}{Basic Books}, \bibinfo{address}{New York, NY}, \bibinfo{pages}{3--32}.
\newblock


\bibitem[Genuis and Bronstein(2017)]%
        {Shelagh}
\bibfield{author}{\bibinfo{person}{Shelagh Genuis} {and} \bibinfo{person}{Jenny Bronstein}.} \bibinfo{year}{2017}\natexlab{}.
\newblock \showarticletitle{Looking for "normal": Sensemaking in the context of health disruption}.
\newblock \bibinfo{journal}{\emph{Journal of the Association for Information Science and Technology}} \bibinfo{volume}{68}, \bibinfo{number}{3} (\bibinfo{year}{2017}), \bibinfo{pages}{750--761}.
\newblock


\bibitem[Gowin et~al\mbox{.}(2019)]%
        {Gowin}
\bibfield{author}{\bibinfo{person}{Mary Gowin}, \bibinfo{person}{Amanda Wilkerson}, \bibinfo{person}{Sarah Maness}, \bibinfo{person}{Daniel~J. Larson}, \bibinfo{person}{H.~Michael Crowson}, \bibinfo{person}{Michael Smith}, {and} \bibinfo{person}{Marshall~K. Cheney}.} \bibinfo{year}{2019}\natexlab{}.
\newblock \showarticletitle{Wearable activity tracker use in young adults through the lens of social cognitive theory}.
\newblock \bibinfo{journal}{\emph{American Journal of Health Education}} \bibinfo{volume}{50}, \bibinfo{number}{1} (\bibinfo{year}{2019}), \bibinfo{pages}{40--51}.
\newblock
\urldef\tempurl%
\url{https://doi.org/10.1080/19325037.2018.1548314}
\showDOI{\tempurl}


\bibitem[Haghayegh et~al\mbox{.}(2091)]%
        {Haghayegh}
\bibfield{author}{\bibinfo{person}{Shahab Haghayegh}, \bibinfo{person}{Sepideh Khoshnevis}, \bibinfo{person}{Michael~H. Smolensky}, \bibinfo{person}{Kenneth~R. Diller}, {and} \bibinfo{person}{Richard~J. Castriotta}.} \bibinfo{year}{2091}\natexlab{}.
\newblock \showarticletitle{Accuracy of wristband Fitbit models in assessing sleep: systematic review and meta-analysis}.
\newblock \bibinfo{journal}{\emph{Journal of medical Internet research}} \bibinfo{volume}{21}, \bibinfo{number}{11} (\bibinfo{year}{2091}), \bibinfo{pages}{e16723}.
\newblock
\urldef\tempurl%
\url{https://doi.org/10.2196/16273}
\showDOI{\tempurl}


\bibitem[Haldar et~al\mbox{.}(2020)]%
        {Haldar20}
\bibfield{author}{\bibinfo{person}{Shefali Haldar}, \bibinfo{person}{Yoojung Kim}, \bibinfo{person}{Sonali~R. Mishra}, \bibinfo{person}{Andrea~L. Hartzler}, \bibinfo{person}{Ari~H. Pollack}, {and} \bibinfo{person}{Wanda Pratt}.} \bibinfo{year}{2020}\natexlab{}.
\newblock \showarticletitle{The Patient Advice System: A Technology Probe Study to Enable Peer Support in the Hospital}.
\newblock \bibinfo{journal}{\emph{Proc. ACM Hum.-Comput. Interact.}} \bibinfo{volume}{4}, \bibinfo{number}{CSCW2}, Article \bibinfo{articleno}{112} (\bibinfo{date}{oct} \bibinfo{year}{2020}), \bibinfo{numpages}{23}~pages.
\newblock
\urldef\tempurl%
\url{https://doi.org/10.1145/3415183}
\showDOI{\tempurl}


\bibitem[Heuschkel and Kauschke(2018)]%
        {Heuschkel}
\bibfield{author}{\bibinfo{person}{Jens Heuschkel} {and} \bibinfo{person}{Sebastian Kauschke}.} \bibinfo{year}{2018}\natexlab{}.
\newblock \showarticletitle{More Data Matters: Improving CGM Prediction via Ubiquitous Data and Deep Learning}. In \bibinfo{booktitle}{\emph{Proceedings of the 2018 ACM International Joint Conference and 2018 International Symposium on Pervasive and Ubiquitous Computing and Wearable Computers}}. \bibinfo{publisher}{Association for Computing Machinery}, \bibinfo{address}{New York, NY, USA}, \bibinfo{pages}{809–816}.
\newblock
\showISBNx{9781450359665}
\urldef\tempurl%
\url{https://doi.org/10.1145/3267305.3274132}
\showDOI{\tempurl}


\bibitem[Homewood(2023)]%
        {Homewood}
\bibfield{author}{\bibinfo{person}{Sarah Homewood}.} \bibinfo{year}{2023}\natexlab{}.
\newblock \showarticletitle{Self-Tracking to Do Less: An Autoethnography of Long COVID That Informs the Design of Pacing Technologies}. In \bibinfo{booktitle}{\emph{Proceedings of the 2023 CHI Conference on Human Factors in Computing Systems}} (Hamburg, Germany) \emph{(\bibinfo{series}{CHI '23})}. \bibinfo{publisher}{Association for Computing Machinery}, \bibinfo{address}{New York, NY, USA}, Article \bibinfo{articleno}{656}, \bibinfo{numpages}{14}~pages.
\newblock
\showISBNx{9781450394215}
\urldef\tempurl%
\url{https://doi.org/10.1145/3544548.3581505}
\showDOI{\tempurl}


\bibitem[Hong et~al\mbox{.}(2020)]%
        {Hong}
\bibfield{author}{\bibinfo{person}{Matthew~K. Hong}, \bibinfo{person}{Udaya Lakshmi}, \bibinfo{person}{Kimberly Do}, \bibinfo{person}{Sampath Prahalad}, \bibinfo{person}{Thomas Olson}, \bibinfo{person}{Rosa~I. Arriaga}, {and} \bibinfo{person}{Lauren Wilcox}.} \bibinfo{year}{2020}\natexlab{}.
\newblock \showarticletitle{Using Diaries to Probe the Illness Experiences of Adolescent Patients and Parental Caregivers}. In \bibinfo{booktitle}{\emph{Proceedings of the 2020 CHI Conference on Human Factors in Computing Systems}} (Honolulu, HI, USA) \emph{(\bibinfo{series}{CHI '20})}. \bibinfo{publisher}{Association for Computing Machinery}, \bibinfo{address}{New York, NY, USA}, \bibinfo{pages}{1–16}.
\newblock
\showISBNx{9781450367080}
\urldef\tempurl%
\url{https://doi.org/10.1145/3313831.3376426}
\showDOI{\tempurl}


\bibitem[Ibáñez et~al\mbox{.}(2018)]%
        {Ibáñez}
\bibfield{author}{\bibinfo{person}{Vanessa Ibáñez}, \bibinfo{person}{Josep Silva}, {and} \bibinfo{person}{Omar Cauli}.} \bibinfo{year}{2018}\natexlab{}.
\newblock \showarticletitle{"A survey on sleep assessment methods}.
\newblock \bibinfo{journal}{\emph{PeerJ}}  \bibinfo{volume}{6} (\bibinfo{year}{2018}), \bibinfo{pages}{e4849}.
\newblock
\urldef\tempurl%
\url{https://doi.org/10.7717/peerj.4849}
\showDOI{\tempurl}


\bibitem[Iott et~al\mbox{.}(2019)]%
        {Iott}
\bibfield{author}{\bibinfo{person}{Bradley Iott}, \bibinfo{person}{Tanner Caverly}, \bibinfo{person}{Astrid Fishstrom}, \bibinfo{person}{Darren King}, \bibinfo{person}{George Meng}, {and} \bibinfo{person}{Allen Flynn}.} \bibinfo{year}{2019}\natexlab{}.
\newblock \showarticletitle{Clinician Perspectives on the User Experience, Configuration, and Scope of Use of a Patient Reported Outcomes (PRO) Dashboard}. In \bibinfo{booktitle}{\emph{Proceedings of the 13th EAI International Conference on Pervasive Computing Technologies for Healthcare}} (Trento, Italy) \emph{(\bibinfo{series}{PervasiveHealth'19})}. \bibinfo{publisher}{Association for Computing Machinery}, \bibinfo{address}{New York, NY, USA}, \bibinfo{pages}{21–30}.
\newblock
\showISBNx{9781450361262}
\urldef\tempurl%
\url{https://doi.org/10.1145/3329189.3329198}
\showDOI{\tempurl}


\bibitem[Jalilian and Khairat(2022)]%
        {Jalilian}
\bibfield{author}{\bibinfo{person}{Laleh Jalilian} {and} \bibinfo{person}{Saif Khairat}.} \bibinfo{year}{2022}\natexlab{}.
\newblock \showarticletitle{The Next-Generation Electronic Health Record in the ICU: A Focus on User-Technology Interface to Optimize Patient Safety and Quality}.
\newblock \bibinfo{journal}{\emph{Perspectives in Health Information Management}} \bibinfo{volume}{19}, \bibinfo{number}{1} (\bibinfo{year}{2022}), \bibinfo{pages}{1g}.
\newblock


\bibitem[Joachim and Acorn(2000)]%
        {Joachim}
\bibfield{author}{\bibinfo{person}{Gloria Joachim} {and} \bibinfo{person}{Sonia Acorn}.} \bibinfo{year}{2000}\natexlab{}.
\newblock \showarticletitle{Stigma of visible and invisible chronic conditions}.
\newblock \bibinfo{journal}{\emph{Journal of advanced nursing}} \bibinfo{volume}{32}, \bibinfo{number}{1} (\bibinfo{year}{2000}), \bibinfo{pages}{243--248}.
\newblock
\urldef\tempurl%
\url{https://doi.org/10.1046/j.1365-2648.2000.01466.x}
\showDOI{\tempurl}


\bibitem[Karaarslan et~al\mbox{.}(2022)]%
        {Karaarslan}
\bibfield{author}{\bibinfo{person}{Fatih Karaarslan}, \bibinfo{person}{Fulya~Demircioğlu Güneri}, {and} \bibinfo{person}{Sinan Kardeş}.} \bibinfo{year}{2022}\natexlab{}.
\newblock \showarticletitle{Long COVID: rheumatologic/musculoskeletal symptoms in hospitalized COVID-19 survivors at 3 and 6 months}.
\newblock \bibinfo{journal}{\emph{Clinical rheumatology}}  \bibinfo{volume}{41} (\bibinfo{year}{2022}), \bibinfo{pages}{289--286}.
\newblock
\urldef\tempurl%
\url{https://doi.org/10.1007/s10067-021-05942-x}
\showDOI{\tempurl}


\bibitem[Karusala and Anderson(2022)]%
        {Karusala}
\bibfield{author}{\bibinfo{person}{Naveena Karusala} {and} \bibinfo{person}{Richard Anderson}.} \bibinfo{year}{2022}\natexlab{}.
\newblock \showarticletitle{Towards Conviviality in NavigatingHealth Information on Social Media}. In \bibinfo{booktitle}{\emph{Proceedings of the 2022 CHI Conference on Human Factors in Computing Systems}} (New Orleans, LA, USA) \emph{(\bibinfo{series}{CHI '22})}. \bibinfo{publisher}{Association for Computing Machinery}, \bibinfo{address}{New York, NY, USA}, Article \bibinfo{articleno}{43}, \bibinfo{numpages}{14}~pages.
\newblock
\showISBNx{9781450391573}
\urldef\tempurl%
\url{https://doi.org/10.1145/3491102.3517622}
\showDOI{\tempurl}


\bibitem[Kawu et~al\mbox{.}(2022)]%
        {Kawu}
\bibfield{author}{\bibinfo{person}{A.A. Kawu}, \bibinfo{person}{L. Hederman}, \bibinfo{person}{D. O'Sullivan}, {and} \bibinfo{person}{J. Doyle}.} \bibinfo{year}{2022}\natexlab{}.
\newblock \showarticletitle{Patient generated health data and electronic health record integration, governance and socio-technical issues: A narrative review}.
\newblock \bibinfo{journal}{\emph{Informatics in Medicine Unlocked}}  \bibinfo{volume}{37} (\bibinfo{year}{2022}), \bibinfo{pages}{101153}.
\newblock
\urldef\tempurl%
\url{https://doi.org/10.1016/j.imu.2022.101153}
\showDOI{\tempurl}


\bibitem[Khondakar and Kaushik(2022)]%
        {Khondakar}
\bibfield{author}{\bibinfo{person}{Kamil~Reza Khondakar} {and} \bibinfo{person}{Ajeet Kaushik}.} \bibinfo{year}{2022}\natexlab{}.
\newblock \showarticletitle{Role of wearable sensing technology to manage long COVID}.
\newblock \bibinfo{journal}{\emph{Biosensors}} \bibinfo{volume}{13}, \bibinfo{number}{1} (\bibinfo{year}{2022}), \bibinfo{pages}{62}.
\newblock
\urldef\tempurl%
\url{https://doi.org/10.3390/bios13010062}
\showDOI{\tempurl}


\bibitem[Kohane et~al\mbox{.}(2021)]%
        {Kohane}
\bibfield{author}{\bibinfo{person}{Isaac~S. Kohane}, \bibinfo{person}{Bruce~J. Aronow}, \bibinfo{person}{Paul Avillach}, \bibinfo{person}{Brett~K. Beaulieu-Jones}, \bibinfo{person}{Riccardo Bellazzi}, \bibinfo{person}{Robert~L. Bradford}, \bibinfo{person}{Gabriel~A. Brat}, \bibinfo{person}{Mario Cannatro}, \bibinfo{person}{James~J Cimino}, \bibinfo{person}{Noelia Garcia-Barrio}, \bibinfo{person}{Nils Gehlenborg}, \bibinfo{person}{Marzyeh Ghassemi}, \bibinfo{person}{Alba Gutierrez-Sacristan}, \bibinfo{person}{David~A Hanauer}, \bibinfo{person}{John~H. Holmes}, \bibinfo{person}{Chuan Hong}, \bibinfo{person}{Jeffrey~G. Klann}, \bibinfo{person}{Ne~Hooi~Will Loh}, {and} \bibinfo{person}{Tianxi Cai}.} \bibinfo{year}{2021}\natexlab{}.
\newblock \showarticletitle{What every reader should know about studies using electronic health record data but may be afraid to ask}.
\newblock \bibinfo{journal}{\emph{Journal of medical Internet research}} \bibinfo{volume}{23}, \bibinfo{number}{3} (\bibinfo{year}{2021}), \bibinfo{pages}{e22219}.
\newblock
\urldef\tempurl%
\url{https://doi.org/10.2196/22219}
\showDOI{\tempurl}


\bibitem[Kooij et~al\mbox{.}(2017)]%
        {Kooij}
\bibfield{author}{\bibinfo{person}{Laura Kooij}, \bibinfo{person}{Wim~G. Groen}, {and} \bibinfo{person}{Wim H.~Van Harten}.} \bibinfo{year}{2017}\natexlab{}.
\newblock \showarticletitle{The effectiveness of information technology-supported shared care for patients with chronic disease: a systematic review}.
\newblock \bibinfo{journal}{\emph{Journal of medical Internet research}} \bibinfo{volume}{19}, \bibinfo{number}{6} (\bibinfo{year}{2017}), \bibinfo{pages}{e7405}.
\newblock
\urldef\tempurl%
\url{https://doi.org/10.2196/jmir.7405}
\showDOI{\tempurl}


\bibitem[Koren and Prasad(2022)]%
        {Koren}
\bibfield{author}{\bibinfo{person}{Ana Koren} {and} \bibinfo{person}{Ramjee Prasad}.} \bibinfo{year}{2022}\natexlab{}.
\newblock \showarticletitle{Iot health data in electronic health records (ehr): Security and privacy issues in era of 6g}.
\newblock \bibinfo{journal}{\emph{Journal of ICT Standardization}} \bibinfo{volume}{10}, \bibinfo{number}{1} (\bibinfo{year}{2022}), \bibinfo{pages}{63--84}.
\newblock
\urldef\tempurl%
\url{https://doi.org/10.13052/jicts2245-800X.1014}
\showDOI{\tempurl}


\bibitem[Last et~al\mbox{.}(2021)]%
        {Last}
\bibfield{author}{\bibinfo{person}{Christina Last}, \bibinfo{person}{Prithviraj Pramanik}, \bibinfo{person}{Nikita Saini}, \bibinfo{person}{Akash~Smaran Majety}, \bibinfo{person}{Do-Hyung Kim}, \bibinfo{person}{Manuel Garc\'{\i}a-Herranz}, {and} \bibinfo{person}{Subhabrata Majumdar}.} \bibinfo{year}{2021}\natexlab{}.
\newblock \showarticletitle{Towards an Open Global Air Quality Monitoring Platform to Assess Children’s Exposure to Air Pollutants in the Light of COVID-19 Lockdowns}. In \bibinfo{booktitle}{\emph{Extended Abstracts of the 2021 CHI Conference on Human Factors in Computing Systems}} (Yokohama, Japan) \emph{(\bibinfo{series}{CHI EA '21})}. \bibinfo{publisher}{Association for Computing Machinery}, \bibinfo{address}{New York, NY, USA}, Article \bibinfo{articleno}{434}, \bibinfo{numpages}{7}~pages.
\newblock
\showISBNx{9781450380959}
\urldef\tempurl%
\url{https://doi.org/10.1145/3411763.3451768}
\showDOI{\tempurl}


\bibitem[Lobach et~al\mbox{.}(2022)]%
        {Lobach}
\bibfield{author}{\bibinfo{person}{David~F. Lobach}, \bibinfo{person}{Aziz Boxwala}, \bibinfo{person}{Nitu Kashyap}, \bibinfo{person}{Krysta Heaney-Huls}, \bibinfo{person}{Andrew~B. Chiao}, \bibinfo{person}{Thomas Rafter}, \bibinfo{person}{Edwin~A. Lomotan}, \bibinfo{person}{Michael~I. Harrison}, \bibinfo{person}{Chris Dymek}, \bibinfo{person}{James Swiger}, {and} \bibinfo{person}{Parshila Dullabh}.} \bibinfo{year}{2022}\natexlab{}.
\newblock \showarticletitle{Integrating a Patient Engagement App into an Electronic Health Record-Enabled Workflow Using Interoperability Standards}.
\newblock \bibinfo{journal}{\emph{Applied Clinical Informatics}} \bibinfo{volume}{13}, \bibinfo{number}{5} (\bibinfo{year}{2022}), \bibinfo{pages}{1163--1171}.
\newblock
\urldef\tempurl%
\url{https://doi.org/10.1055/s-0042-1758736}
\showDOI{\tempurl}


\bibitem[Lv et~al\mbox{.}(2017)]%
        {Lv}
\bibfield{author}{\bibinfo{person}{Nan Lv}, \bibinfo{person}{Lan Xiao}, \bibinfo{person}{Martha~L. Simmons}, \bibinfo{person}{Lisa~G. Rosas}, \bibinfo{person}{Albert Chan}, {and} \bibinfo{person}{Martin Entwistle}.} \bibinfo{year}{2017}\natexlab{}.
\newblock \showarticletitle{Personalized hypertension management using patient-generated health data integrated with electronic health records (EMPOWER-H): six-month pre-post study}.
\newblock \bibinfo{journal}{\emph{Journal of medical Internet research}} \bibinfo{volume}{19}, \bibinfo{number}{9} (\bibinfo{year}{2017}), \bibinfo{pages}{e311}.
\newblock
\urldef\tempurl%
\url{https://doi.org/10.2196/jmir.7831}
\showDOI{\tempurl}


\bibitem[MacLeod et~al\mbox{.}(2015)]%
        {MacLeod}
\bibfield{author}{\bibinfo{person}{Haley MacLeod}, \bibinfo{person}{Kim Oakes}, \bibinfo{person}{Danika Geisler}, \bibinfo{person}{Kay Connelly}, {and} \bibinfo{person}{Katie Siek}.} \bibinfo{year}{2015}\natexlab{}.
\newblock \showarticletitle{Rare World: Towards Technology for Rare Diseases}. In \bibinfo{booktitle}{\emph{Proceedings of the 33rd Annual ACM Conference on Human Factors in Computing Systems}} (Seoul, Republic of Korea) \emph{(\bibinfo{series}{CHI '15})}. \bibinfo{publisher}{Association for Computing Machinery}, \bibinfo{address}{New York, NY, USA}, \bibinfo{pages}{1145–1154}.
\newblock
\showISBNx{9781450331456}
\urldef\tempurl%
\url{https://doi.org/10.1145/2702123.2702494}
\showDOI{\tempurl}


\bibitem[Maestre et~al\mbox{.}(2018)]%
        {Maestre}
\bibfield{author}{\bibinfo{person}{Juan~F. Maestre}, \bibinfo{person}{Haley MacLeod}, \bibinfo{person}{Ciabhan~L. Connelly}, \bibinfo{person}{Julia~C. Dunbar}, \bibinfo{person}{Jordan Beck}, \bibinfo{person}{Katie~A. Siek}, {and} \bibinfo{person}{Patrick~C. Shih}.} \bibinfo{year}{2018}\natexlab{}.
\newblock \showarticletitle{Defining Through Expansion: Conducting Asynchronous Remote Communities (ARC) Research with Stigmatized Groups}. In \bibinfo{booktitle}{\emph{Proceedings of the 2018 CHI Conference on Human Factors in Computing Systems}} (, Montreal QC, Canada,) \emph{(\bibinfo{series}{CHI '18})}. \bibinfo{publisher}{Association for Computing Machinery}, \bibinfo{address}{New York, NY, USA}, \bibinfo{pages}{1–13}.
\newblock
\showISBNx{9781450356206}
\urldef\tempurl%
\url{https://doi.org/10.1145/3173574.3174131}
\showDOI{\tempurl}


\bibitem[Mamykina et~al\mbox{.}(2008)]%
        {Mamykina}
\bibfield{author}{\bibinfo{person}{Lena Mamykina}, \bibinfo{person}{Elizabeth Mynatt}, \bibinfo{person}{Patricia Davidson}, {and} \bibinfo{person}{Daniel Greenblatt}.} \bibinfo{year}{2008}\natexlab{}.
\newblock \showarticletitle{MAHI: Investigation of Social Scaffolding for Reflective Thinking in Diabetes Management}. In \bibinfo{booktitle}{\emph{Proceedings of the SIGCHI Conference on Human Factors in Computing Systems}} (Florence, Italy) \emph{(\bibinfo{series}{CHI '08})}. \bibinfo{publisher}{Association for Computing Machinery}, \bibinfo{address}{New York, NY, USA}, \bibinfo{pages}{477–486}.
\newblock
\showISBNx{9781605580111}
\urldef\tempurl%
\url{https://doi.org/10.1145/1357054.1357131}
\showDOI{\tempurl}


\bibitem[Mankoff et~al\mbox{.}(2011)]%
        {Mankoff}
\bibfield{author}{\bibinfo{person}{Jennifer Mankoff}, \bibinfo{person}{Kateryna Kuksenok}, \bibinfo{person}{Sara Kiesler}, \bibinfo{person}{Jennifer~A. Rode}, {and} \bibinfo{person}{Kelly Waldman}.} \bibinfo{year}{2011}\natexlab{}.
\newblock \showarticletitle{Competing Online Viewpoints and Models of Chronic Illness}. In \bibinfo{booktitle}{\emph{Proceedings of the SIGCHI Conference on Human Factors in Computing Systems}} (Vancouver, BC, Canada) \emph{(\bibinfo{series}{CHI '11})}. \bibinfo{publisher}{Association for Computing Machinery}, \bibinfo{address}{New York, NY, USA}, \bibinfo{pages}{589–598}.
\newblock
\showISBNx{9781450302289}
\urldef\tempurl%
\url{https://doi.org/10.1145/1978942.1979027}
\showDOI{\tempurl}


\bibitem[Marathe et~al\mbox{.}(2021)]%
        {Marthe21}
\bibfield{author}{\bibinfo{person}{Megh Marathe}, \bibinfo{person}{Yoonseon Yi}, \bibinfo{person}{Chia-Hsuan Su}, \bibinfo{person}{Ting-Wei Chang}, {and} \bibinfo{person}{Gabriela Marcu}.} \bibinfo{year}{2021}\natexlab{}.
\newblock \showarticletitle{Tedious Versus Taxing: The Nature of Work in a Behavioral Health Context}.
\newblock \bibinfo{journal}{\emph{Proc. ACM Hum.-Comput. Interact.}} \bibinfo{volume}{5}, \bibinfo{number}{CSCW2}, Article \bibinfo{articleno}{302} (\bibinfo{date}{oct} \bibinfo{year}{2021}), \bibinfo{numpages}{24}~pages.
\newblock
\urldef\tempurl%
\url{https://doi.org/10.1145/3476043}
\showDOI{\tempurl}


\bibitem[Marx(2021)]%
        {Marx}
\bibfield{author}{\bibinfo{person}{Viven Marx}.} \bibinfo{year}{2021}\natexlab{}.
\newblock \showarticletitle{Scientists set out to connect the dots on long COVID." Nature Methods}.
\newblock \bibinfo{journal}{\emph{Nature Methods}} \bibinfo{volume}{18}, \bibinfo{number}{5} (\bibinfo{year}{2021}), \bibinfo{pages}{449--453}.
\newblock
\urldef\tempurl%
\url{https://doi.org/10.1038/s41592-021-01145-z}
\showDOI{\tempurl}


\bibitem[Mead(2022)]%
        {Mead}
\bibfield{author}{\bibinfo{person}{Jessica Mead}.} \bibinfo{year}{2022}\natexlab{}.
\newblock \showarticletitle{Investigation into a Self-Management App for People with Postural Tachycardia Syndrome (POTS)}. In \bibinfo{booktitle}{\emph{Extended Abstracts of the 2022 CHI Conference on Human Factors in Computing Systems}} (New Orleans, LA, USA) \emph{(\bibinfo{series}{CHI EA '22})}. \bibinfo{publisher}{Association for Computing Machinery}, \bibinfo{address}{New York, NY, USA}, Article \bibinfo{articleno}{470}, \bibinfo{numpages}{5}~pages.
\newblock
\showISBNx{9781450391566}
\urldef\tempurl%
\url{https://doi.org/10.1145/3491101.3516803}
\showDOI{\tempurl}


\bibitem[Mehandru and Merad(2022)]%
        {Mehandru}
\bibfield{author}{\bibinfo{person}{Saurabh Mehandru} {and} \bibinfo{person}{Miriam Merad}.} \bibinfo{year}{2022}\natexlab{}.
\newblock \showarticletitle{Pathological sequelae of long-haul COVID}.
\newblock \bibinfo{journal}{\emph{Nature immunology}} \bibinfo{volume}{23}, \bibinfo{number}{2} (\bibinfo{year}{2022}), \bibinfo{pages}{194--202}.
\newblock
\urldef\tempurl%
\url{https://doi.org/10.1038/s41590-021-01104-y}
\showDOI{\tempurl}


\bibitem[Mills et~al\mbox{.}(2007)]%
        {Mills07}
\bibfield{author}{\bibinfo{person}{P.J. Mills}, \bibinfo{person}{R.~Von Kanel}, \bibinfo{person}{D. Norman}, \bibinfo{person}{L. Natarajan}, \bibinfo{person}{M.~G. Ziegler}, {and} \bibinfo{person}{J.E. Dimsdale}.} \bibinfo{year}{2007}\natexlab{}.
\newblock \showarticletitle{Inflammation and sleep in healthy individuals}.
\newblock \bibinfo{journal}{\emph{Sleep}} \bibinfo{volume}{30}, \bibinfo{number}{6} (\bibinfo{year}{2007}), \bibinfo{pages}{729--735}.
\newblock


\bibitem[Moorhead et~al\mbox{.}(2013)]%
        {Moorhead}
\bibfield{author}{\bibinfo{person}{S.~Anne Moorhead}, \bibinfo{person}{Diane~E. Hazlett}, \bibinfo{person}{Laura Harrison}, \bibinfo{person}{Jennifer~K. Carroll}, \bibinfo{person}{Anthea Irwin}, {and} \bibinfo{person}{Ciska Hoving}.} \bibinfo{year}{2013}\natexlab{}.
\newblock \showarticletitle{A new dimension of health care: systematic review of the uses, benefits, and limitations of social media for health communication}.
\newblock \bibinfo{journal}{\emph{Journal of medical Internet research}} \bibinfo{volume}{15}, \bibinfo{number}{4} (\bibinfo{year}{2013}), \bibinfo{pages}{e1933}.
\newblock
\urldef\tempurl%
\url{https://doi.org/10.2196/jmir.1933}
\showDOI{\tempurl}


\bibitem[Morse(2009)]%
        {Morse}
\bibfield{author}{\bibinfo{person}{J.M. Morse}.} \bibinfo{year}{2009}\natexlab{}.
\newblock \showarticletitle{Mixing qualitative methods}.
\newblock \bibinfo{journal}{\emph{Qualitative Health Research}}  \bibinfo{volume}{19} (\bibinfo{year}{2009}), \bibinfo{pages}{15231524}.
\newblock
\urldef\tempurl%
\url{https://doi.org/10.1177/1049732309349360}
\showDOI{\tempurl}


\bibitem[Mosaly et~al\mbox{.}(2016)]%
        {Marks16}
\bibfield{author}{\bibinfo{person}{Prithima~Reddy Mosaly}, \bibinfo{person}{Lukasz Mazur}, {and} \bibinfo{person}{Lawrence~B. Marks}.} \bibinfo{year}{2016}\natexlab{}.
\newblock \showarticletitle{Usability Evaluation of Electronic Health Record System (EHRs) Using Subjective and Objective Measures}. In \bibinfo{booktitle}{\emph{Proceedings of the 2016 ACM on Conference on Human Information Interaction and Retrieval}} (Carrboro, North Carolina, USA) \emph{(\bibinfo{series}{CHIIR '16})}. \bibinfo{publisher}{Association for Computing Machinery}, \bibinfo{address}{New York, NY, USA}, \bibinfo{pages}{313–316}.
\newblock
\showISBNx{9781450337519}
\urldef\tempurl%
\url{https://doi.org/10.1145/2854946.2854985}
\showDOI{\tempurl}


\bibitem[Murnane et~al\mbox{.}(2018)]%
        {Murnane}
\bibfield{author}{\bibinfo{person}{Elizabeth~L. Murnane}, \bibinfo{person}{Tara~G. Walker}, \bibinfo{person}{Beck Tench}, \bibinfo{person}{Stephen Voida}, {and} \bibinfo{person}{Jaime Snyder}.} \bibinfo{year}{2018}\natexlab{}.
\newblock \showarticletitle{Personal Informatics in Interpersonal Contexts: Towards the Design of Technology That Supports the Social Ecologies of Long-Term Mental Health Management}.
\newblock \bibinfo{journal}{\emph{Proc. ACM Hum.-Comput. Interact.}} \bibinfo{volume}{2}, \bibinfo{number}{CSCW}, Article \bibinfo{articleno}{127} (\bibinfo{date}{nov} \bibinfo{year}{2018}), \bibinfo{numpages}{27}~pages.
\newblock
\urldef\tempurl%
\url{https://doi.org/10.1145/3274396}
\showDOI{\tempurl}


\bibitem[Murphy and Reddy(2017)]%
        {Murphy17}
\bibfield{author}{\bibinfo{person}{Alison~R. Murphy} {and} \bibinfo{person}{Madhu~C. Reddy}.} \bibinfo{year}{2017}\natexlab{}.
\newblock \showarticletitle{Ambiguous Accountability: The Challenges of Identifying and Managing Patient-Related Information Problems in Collaborative Patient-Care Teams}. In \bibinfo{booktitle}{\emph{Proceedings of the 2017 ACM Conference on Computer Supported Cooperative Work and Social Computing}} \emph{(\bibinfo{series}{CSCW '17})}. \bibinfo{publisher}{Association for Computing Machinery}, \bibinfo{address}{New York, NY, USA}, \bibinfo{pages}{1646–1660}.
\newblock
\showISBNx{9781450343350}
\urldef\tempurl%
\url{https://doi.org/10.1145/2998181.2998315}
\showDOI{\tempurl}


\bibitem[Oh et~al\mbox{.}(2022)]%
        {Oh22}
\bibfield{author}{\bibinfo{person}{Chi~Young Oh}, \bibinfo{person}{Yuhan Luo}, \bibinfo{person}{Beth St.~Jean}, {and} \bibinfo{person}{Eun~Kyoung Choe}.} \bibinfo{year}{2022}\natexlab{}.
\newblock \showarticletitle{Patients Waiting for Cues: Information Asymmetries and Challenges in Sharing Patient-Generated Data in the Clinic}.
\newblock \bibinfo{journal}{\emph{Proc. ACM Hum.-Comput. Interact.}} \bibinfo{volume}{6}, \bibinfo{number}{CSCW1}, Article \bibinfo{articleno}{107} (\bibinfo{date}{apr} \bibinfo{year}{2022}), \bibinfo{numpages}{23}~pages.
\newblock
\urldef\tempurl%
\url{https://doi.org/10.1145/3512954}
\showDOI{\tempurl}


\bibitem[O'Kane and Mentis(2012)]%
        {O'Kane}
\bibfield{author}{\bibinfo{person}{Aisling~Ann O'Kane} {and} \bibinfo{person}{Helena Mentis}.} \bibinfo{year}{2012}\natexlab{}.
\newblock \showarticletitle{Sharing Medical Data vs. Health Knowledge in Chronic Illness Care}. In \bibinfo{booktitle}{\emph{CHI '12 Extended Abstracts on Human Factors in Computing Systems}} (Austin, Texas, USA) \emph{(\bibinfo{series}{CHI EA '12})}. \bibinfo{publisher}{Association for Computing Machinery}, \bibinfo{address}{New York, NY, USA}, \bibinfo{pages}{2417–2422}.
\newblock
\showISBNx{9781450310161}
\urldef\tempurl%
\url{https://doi.org/10.1145/2212776.2223812}
\showDOI{\tempurl}


\bibitem[Orhun and Yıldırım(2021)]%
        {Orhun}
\bibfield{author}{\bibinfo{person}{S.E. Orhun} {and} \bibinfo{person}{Y. Yıldırım}.} \bibinfo{year}{2021}\natexlab{}.
\newblock \showarticletitle{Invisible to Visible: Identifying the Emerging Communication Needs in the ‘New Normal’through Design Research}.
\newblock \bibinfo{journal}{\emph{Strategic Design Research Journal}} \bibinfo{volume}{14}, \bibinfo{number}{1} (\bibinfo{year}{2021}), \bibinfo{pages}{1--13}.
\newblock
\urldef\tempurl%
\url{https://doi.org/10.4013/sdrj.2021.14.21408}
\showDOI{\tempurl}


\bibitem[Pang et~al\mbox{.}(2013)]%
        {Pang}
\bibfield{author}{\bibinfo{person}{Carolyn~E. Pang}, \bibinfo{person}{Carman Neustaedter}, \bibinfo{person}{Bernhard~E. Riecke}, \bibinfo{person}{Erick Oduor}, {and} \bibinfo{person}{Serena Hillman}.} \bibinfo{year}{2013}\natexlab{}.
\newblock \showarticletitle{Technology Preferences and Routines for Sharing Health Information during the Treatment of a Chronic Illness}. In \bibinfo{booktitle}{\emph{Proceedings of the SIGCHI Conference on Human Factors in Computing Systems}} (Paris, France) \emph{(\bibinfo{series}{CHI '13})}. \bibinfo{publisher}{Association for Computing Machinery}, \bibinfo{address}{New York, NY, USA}, \bibinfo{pages}{1759–1768}.
\newblock
\showISBNx{9781450318990}
\urldef\tempurl%
\url{https://doi.org/10.1145/2470654.2466232}
\showDOI{\tempurl}


\bibitem[Patel et~al\mbox{.}(2019)]%
        {Patel}
\bibfield{author}{\bibinfo{person}{Dilisha Patel}, \bibinfo{person}{Ann Blandford}, \bibinfo{person}{Mark Warner}, \bibinfo{person}{Jill Shawe}, {and} \bibinfo{person}{Judith Stephenson}.} \bibinfo{year}{2019}\natexlab{}.
\newblock \showarticletitle{"I Feel like Only Half a Man": Online Forums as a Resource for Finding a "New Normal" for Men Experiencing Fertility Issues}.
\newblock \bibinfo{journal}{\emph{Proc. ACM Hum.-Comput. Interact.}} \bibinfo{volume}{3}, \bibinfo{number}{CSCW}, Article \bibinfo{articleno}{82} (\bibinfo{date}{nov} \bibinfo{year}{2019}), \bibinfo{numpages}{20}~pages.
\newblock
\urldef\tempurl%
\url{https://doi.org/10.1145/3359184}
\showDOI{\tempurl}


\bibitem[Pater et~al\mbox{.}(2023)]%
        {Pater23}
\bibfield{author}{\bibinfo{person}{Jessica~A Pater}, \bibinfo{person}{Amanda Coupe}, \bibinfo{person}{Fayika~Farhat Nova}, \bibinfo{person}{Rachel Pfafman}, \bibinfo{person}{Jeanne Carroll}, \bibinfo{person}{Abigal Brouwer}, \bibinfo{person}{Camden Bohn}, \bibinfo{person}{Jason Li}, \bibinfo{person}{Noah Todd}, \bibinfo{person}{Fen~Lei Chang}, {and} \bibinfo{person}{Shion Guha}.} \bibinfo{year}{2023}\natexlab{}.
\newblock \showarticletitle{Social Media is Not a Health Proxy: Differences Between Social Media and Electronic Health Record Reports of Post-COVID Symptoms}.
\newblock \bibinfo{journal}{\emph{Proc. ACM Hum.-Comput. Interact.}} \bibinfo{volume}{7}, \bibinfo{number}{CSCW1}, Article \bibinfo{articleno}{148} (\bibinfo{date}{apr} \bibinfo{year}{2023}), \bibinfo{numpages}{25}~pages.
\newblock
\urldef\tempurl%
\url{https://doi.org/10.1145/3579624}
\showDOI{\tempurl}


\bibitem[Pater et~al\mbox{.}(2019a)]%
        {Pater2019}
\bibfield{author}{\bibinfo{person}{Jessica~A. Pater}, \bibinfo{person}{Brooke Farrington}, \bibinfo{person}{Alycia Brown}, \bibinfo{person}{Lauren~E. Reining}, \bibinfo{person}{Tammy Toscos}, {and} \bibinfo{person}{Elizabeth~D. Mynatt}.} \bibinfo{year}{2019}\natexlab{a}.
\newblock \showarticletitle{Exploring Indicators of Digital Self-Harm with Eating Disorder Patients: A Case Study}.
\newblock \bibinfo{journal}{\emph{Proc. ACM Hum.-Comput. Interact.}} \bibinfo{volume}{3}, \bibinfo{number}{CSCW}, Article \bibinfo{articleno}{84} (\bibinfo{date}{nov} \bibinfo{year}{2019}), \bibinfo{numpages}{26}~pages.
\newblock
\urldef\tempurl%
\url{https://doi.org/10.1145/3359186}
\showDOI{\tempurl}


\bibitem[Pater et~al\mbox{.}(2019b)]%
        {Pater19}
\bibfield{author}{\bibinfo{person}{Jessica~A. Pater}, \bibinfo{person}{Brooke Farrington}, \bibinfo{person}{Alycia Brown}, \bibinfo{person}{Lauren~E. Reining}, \bibinfo{person}{Tammy Toscos}, {and} \bibinfo{person}{Elizabeth~D. Mynatt}.} \bibinfo{year}{2019}\natexlab{b}.
\newblock \showarticletitle{Exploring Indicators of Digital Self-Harm with Eating Disorder Patients: A Case Study}.
\newblock \bibinfo{journal}{\emph{Proc. ACM Hum.-Comput. Interact.}} \bibinfo{volume}{3}, \bibinfo{number}{CSCW}, Article \bibinfo{articleno}{84} (\bibinfo{date}{nov} \bibinfo{year}{2019}), \bibinfo{numpages}{26}~pages.
\newblock
\urldef\tempurl%
\url{https://doi.org/10.1145/3359186}
\showDOI{\tempurl}


\bibitem[Pater et~al\mbox{.}(2021)]%
        {Pater21}
\bibfield{author}{\bibinfo{person}{Jessica~A. Pater}, \bibinfo{person}{Chanda Phelan}, \bibinfo{person}{Victor~P. Cornet}, \bibinfo{person}{Ryan Ahmed}, \bibinfo{person}{Sarah Colletta}, \bibinfo{person}{Erik Hess}, \bibinfo{person}{Connie Kerrigan}, {and} \bibinfo{person}{Tammy Toscos}.} \bibinfo{year}{2021}\natexlab{}.
\newblock \showarticletitle{User-Centered Design of a Mobile App to Support Peer Recovery in a Clinical Setting}.
\newblock \bibinfo{journal}{\emph{Proc. ACM Hum.-Comput. Interact.}} \bibinfo{volume}{5}, \bibinfo{number}{CSCW1}, Article \bibinfo{articleno}{112} (\bibinfo{date}{apr} \bibinfo{year}{2021}), \bibinfo{numpages}{31}~pages.
\newblock
\urldef\tempurl%
\url{https://doi.org/10.1145/3449186}
\showDOI{\tempurl}


\bibitem[Patton(1999)]%
        {Patton}
\bibfield{author}{\bibinfo{person}{M.Q. Patton}.} \bibinfo{year}{1999}\natexlab{}.
\newblock \showarticletitle{Enhancing the quality and credibility of qualitative analysis}.
\newblock \bibinfo{journal}{\emph{Health Sciences Research}}  \bibinfo{volume}{34} (\bibinfo{year}{1999}), \bibinfo{pages}{1189--–1208}.
\newblock


\bibitem[Phillips and Williams(2021)]%
        {Phillips}
\bibfield{author}{\bibinfo{person}{Steven Phillips} {and} \bibinfo{person}{Michelle~A. Williams}.} \bibinfo{year}{2021}\natexlab{}.
\newblock \showarticletitle{Confronting our next national health disaster—long-haul Covid}.
\newblock \bibinfo{journal}{\emph{New England Journal of Medicine}} \bibinfo{volume}{385}, \bibinfo{number}{7} (\bibinfo{year}{2021}), \bibinfo{pages}{577--579}.
\newblock
\urldef\tempurl%
\url{https://doi.org/10.1056/NEJMp2109285}
\showDOI{\tempurl}


\bibitem[Plastiras and O’Sullivan(2018)]%
        {Plastiras}
\bibfield{author}{\bibinfo{person}{Panagiotis Plastiras} {and} \bibinfo{person}{Dympna O’Sullivan}.} \bibinfo{year}{2018}\natexlab{}.
\newblock \showarticletitle{Exchanging personal health data with electronic health records: A standardized information model for patient generated health data and observations of daily living}.
\newblock \bibinfo{journal}{\emph{International journal of medical informatics}}  \bibinfo{volume}{120} (\bibinfo{year}{2018}), \bibinfo{pages}{116--125}.
\newblock
\urldef\tempurl%
\url{https://doi.org/10.1016/j.ijmedinf.2018.10.006}
\showDOI{\tempurl}


\bibitem[Rao and Joshi(2020)]%
        {Rao}
\bibfield{author}{\bibinfo{person}{Pallavi Rao} {and} \bibinfo{person}{Anirudha Joshi}.} \bibinfo{year}{2020}\natexlab{}.
\newblock \showarticletitle{Design Opportunities for Supporting Elderly in India in Managing Their Health and Fitness Post-COVID-19}. In \bibinfo{booktitle}{\emph{IndiaHCI '20: Proceedings of the 11th Indian Conference on Human-Computer Interaction}} (Online, India) \emph{(\bibinfo{series}{IndiaHCI 2020})}. \bibinfo{publisher}{Association for Computing Machinery}, \bibinfo{address}{New York, NY, USA}, \bibinfo{pages}{34–41}.
\newblock
\showISBNx{9781450389440}
\urldef\tempurl%
\url{https://doi.org/10.1145/3429290.3429294}
\showDOI{\tempurl}


\bibitem[Raths(2022)]%
        {Raths}
\bibfield{author}{\bibinfo{person}{David Raths}.} \bibinfo{year}{2022}\natexlab{}.
\newblock \bibinfo{title}{Regenstrief Developing Data Infrastructure to Study Long COVID}.
\newblock
\newblock
\urldef\tempurl%
\url{https://www.hcinnovationgroup.com/policy-value-based-care/public-health/news/21280104/regenstrief-developing-data-infrastructure-to-study-long-covid}
\showURL{%
\tempurl}


\bibitem[Roche et~al\mbox{.}(2017)]%
        {Roche}
\bibfield{author}{\bibinfo{person}{Lauren Roche}, \bibinfo{person}{David~L. Dawson}, \bibinfo{person}{Nima~G. Moghaddam}, \bibinfo{person}{Anne Abey}, {and} \bibinfo{person}{David~M. Gresswell}.} \bibinfo{year}{2017}\natexlab{}.
\newblock \showarticletitle{An Acceptance and Commitment Therapy (ACT) intervention for Chronic Fatigue Syndrome (CFS): a case series approach}.
\newblock \bibinfo{journal}{\emph{Journal of Contextual Behavioral Science}} \bibinfo{volume}{6}, \bibinfo{number}{2} (\bibinfo{year}{2017}), \bibinfo{pages}{178--186}.
\newblock
\urldef\tempurl%
\url{https://doi.org/10.1016/j.jcbs.2017.04.007}
\showDOI{\tempurl}


\bibitem[Rogers et~al\mbox{.}(2010)]%
        {Rogers}
\bibfield{author}{\bibinfo{person}{Yvonne Rogers}, \bibinfo{person}{William~R. Hazlewood}, \bibinfo{person}{Paul Marshall}, \bibinfo{person}{Nick Dalton}, {and} \bibinfo{person}{Susanna Hertrich}.} \bibinfo{year}{2010}\natexlab{}.
\newblock \showarticletitle{Ambient Influence: Can Twinkly Lights Lure and Abstract Representations Trigger Behavioral Change?}. In \bibinfo{booktitle}{\emph{Proceedings of the 12th ACM International Conference on Ubiquitous Computing}} (Copenhagen, Denmark) \emph{(\bibinfo{series}{UbiComp '10})}. \bibinfo{publisher}{Association for Computing Machinery}, \bibinfo{address}{New York, NY, USA}, \bibinfo{pages}{261–270}.
\newblock
\showISBNx{9781605588438}
\urldef\tempurl%
\url{https://doi.org/10.1145/1864349.1864372}
\showDOI{\tempurl}


\bibitem[Rooksby et~al\mbox{.}(2014)]%
        {Rooksby}
\bibfield{author}{\bibinfo{person}{John Rooksby}, \bibinfo{person}{Mattias Rost}, \bibinfo{person}{Alistair Morrison}, {and} \bibinfo{person}{Matthew Chalmers}.} \bibinfo{year}{2014}\natexlab{}.
\newblock \showarticletitle{Personal Tracking as Lived Informatics}. In \bibinfo{booktitle}{\emph{Proceedings of the SIGCHI Conference on Human Factors in Computing Systems}} (Toronto, Ontario, Canada) \emph{(\bibinfo{series}{CHI '14})}. \bibinfo{publisher}{Association for Computing Machinery}, \bibinfo{address}{New York, NY, USA}, \bibinfo{pages}{1163–1172}.
\newblock
\showISBNx{9781450324731}
\urldef\tempurl%
\url{https://doi.org/10.1145/2556288.2557039}
\showDOI{\tempurl}


\bibitem[Rosenberg et~al\mbox{.}(2016)]%
        {Rosenberg}
\bibfield{author}{\bibinfo{person}{Dori Rosenberg}, \bibinfo{person}{Elyse~A. Kadokura}, \bibinfo{person}{Erin~D. Bouldin}, \bibinfo{person}{Christina~E. Miyawaki}, \bibinfo{person}{Celestia~S. Higano}, {and} \bibinfo{person}{Andrea~L. Hartzler}.} \bibinfo{year}{2016}\natexlab{}.
\newblock \showarticletitle{Acceptability of Fitbit for physical activity tracking within clinical care among men with prostate cancer}. In \bibinfo{booktitle}{\emph{AMIA annual symposium proceedings}} (Washington D.C.) \emph{(\bibinfo{series}{AMIA '16})}. \bibinfo{publisher}{American Medical Informatics Association}, \bibinfo{address}{Washington D.C.}, \bibinfo{pages}{1050}.
\newblock


\bibitem[Rossi et~al\mbox{.}(2018)]%
        {Rossi}
\bibfield{author}{\bibinfo{person}{Amerigo Rossi}, \bibinfo{person}{Laena Frechette}, \bibinfo{person}{Devin Miller}, \bibinfo{person}{Eirwen Miller}, \bibinfo{person}{Ciaran Friel}, \bibinfo{person}{Anne~Van Arsdale}, \bibinfo{person}{Juan Lin}, \bibinfo{person}{Viswanathan Shankar}, \bibinfo{person}{Dennis~YS Kuo}, {and} \bibinfo{person}{Nicole~S. Nevadunsky}.} \bibinfo{year}{2018}\natexlab{}.
\newblock \showarticletitle{Acceptability and feasibility of a Fitbit physical activity monitor for endometrial cancer survivors}.
\newblock \bibinfo{journal}{\emph{Gynecologic oncology}} \bibinfo{volume}{149}, \bibinfo{number}{3} (\bibinfo{year}{2018}), \bibinfo{pages}{470--475}.
\newblock
\urldef\tempurl%
\url{https://doi.org/10.1016/j.ygyno.2018.04.560}
\showDOI{\tempurl}


\bibitem[Salamah et~al\mbox{.}(2021)]%
        {Salamah}
\bibfield{author}{\bibinfo{person}{Yasmin Salamah}, \bibinfo{person}{Rahma~Dany Asyifa}, {and} \bibinfo{person}{Auzi Asfarian}.} \bibinfo{year}{2021}\natexlab{}.
\newblock \showarticletitle{Improving The Usability of Personal Health Record in Mobile Health Application for People with Autoimmune Disease}. In \bibinfo{booktitle}{\emph{Asian CHI Symposium 2021}} (Yokohama, Japan) \emph{(\bibinfo{series}{Asian CHI Symposium 2021})}. \bibinfo{publisher}{Association for Computing Machinery}, \bibinfo{address}{New York, NY, USA}, \bibinfo{pages}{180–188}.
\newblock
\showISBNx{9781450382038}
\urldef\tempurl%
\url{https://doi.org/10.1145/3429360.3468207}
\showDOI{\tempurl}


\bibitem[Shi(2021)]%
        {Shi}
\bibfield{author}{\bibinfo{person}{Jiayue Shi}.} \bibinfo{year}{2021}\natexlab{}.
\newblock \showarticletitle{Contributions of Blended Teaching Mode to Interdisciplinary Education in the Post-Covid-19 Era}. In \bibinfo{booktitle}{\emph{2021 5th International Conference on Education and E-Learning}} (Virtual Event, Japan) \emph{(\bibinfo{series}{ICEEL 2021})}. \bibinfo{publisher}{Association for Computing Machinery}, \bibinfo{address}{New York, NY, USA}, \bibinfo{pages}{129–133}.
\newblock
\showISBNx{9781450385749}
\urldef\tempurl%
\url{https://doi.org/10.1145/3502434.3502475}
\showDOI{\tempurl}


\bibitem[Siek(2011)]%
        {Siek}
\bibfield{author}{\bibinfo{person}{Katie~A. Siek}.} \bibinfo{year}{2011}\natexlab{}.
\newblock \showarticletitle{What Are Our Responsibilities When Designing Sociotechnical Health Interventions?}
\newblock \bibinfo{journal}{\emph{Interactions}} \bibinfo{volume}{18}, \bibinfo{number}{5} (\bibinfo{date}{sep} \bibinfo{year}{2011}), \bibinfo{pages}{20–23}.
\newblock
\showISSN{1072-5520}
\urldef\tempurl%
\url{https://doi.org/10.1145/2008176.2008183}
\showDOI{\tempurl}


\bibitem[Soriano et~al\mbox{.}(2022)]%
        {Soriano}
\bibfield{author}{\bibinfo{person}{JB. Soriano}, \bibinfo{person}{S. Murthy}, \bibinfo{person}{J.C. Marshall}, \bibinfo{person}{P. Relan}, {and} \bibinfo{person}{J.V. Diaz}.} \bibinfo{year}{2022}\natexlab{}.
\newblock \showarticletitle{WHO Clinical Case Definition Working Group on Post-COVID-19 Condition. A clinical case definition of post-COVID-19 condition by a Delphi consensus}.
\newblock \bibinfo{journal}{\emph{Lancet Infectious Disease}} \bibinfo{volume}{22}, \bibinfo{number}{4} (\bibinfo{year}{2022}), \bibinfo{pages}{e102--e107}.
\newblock
\urldef\tempurl%
\url{https://doi.org/10.1016/S1473-3099(21)00703-9}
\showDOI{\tempurl}


\bibitem[Staff(2020)]%
        {mayo}
\bibfield{author}{\bibinfo{person}{Mayo~Clinic Staff}.} \bibinfo{year}{2020}\natexlab{}.
\newblock \bibinfo{title}{10,000 steps a day: Too low? Too high?}
\newblock
\newblock
\urldef\tempurl%
\url{https://www.mayoclinic.org/healthy-lifestyle/fitness/in-depth/10000-steps/art-20317391}
\showURL{%
\tempurl}


\bibitem[Sun et~al\mbox{.}(2013)]%
        {Sun}
\bibfield{author}{\bibinfo{person}{Si Sun}, \bibinfo{person}{Xiaomu Zhou}, \bibinfo{person}{Joshua~C. Denny}, \bibinfo{person}{Trent~S. Rosenbloom}, {and} \bibinfo{person}{Hua Xu}.} \bibinfo{year}{2013}\natexlab{}.
\newblock \showarticletitle{Messaging to Your Doctors: Understanding Patient-Provider Communications via a Portal System}. In \bibinfo{booktitle}{\emph{Proceedings of the SIGCHI Conference on Human Factors in Computing Systems}} (Paris, France) \emph{(\bibinfo{series}{CHI '13})}. \bibinfo{publisher}{Association for Computing Machinery}, \bibinfo{address}{New York, NY, USA}, \bibinfo{pages}{1739–1748}.
\newblock
\showISBNx{9781450318990}
\urldef\tempurl%
\url{https://doi.org/10.1145/2470654.2466230}
\showDOI{\tempurl}


\bibitem[Tang et~al\mbox{.}(2015)]%
        {Tang15}
\bibfield{author}{\bibinfo{person}{Charlotte Tang}, \bibinfo{person}{Yunan Chen}, \bibinfo{person}{Bryan~C. Semaan}, {and} \bibinfo{person}{Jahmeilah~A. Roberson}.} \bibinfo{year}{2015}\natexlab{}.
\newblock \showarticletitle{Restructuring Human Infrastructure: The Impact of EHR Deployment in a Volunteer-Dependent Clinic}. In \bibinfo{booktitle}{\emph{Proceedings of the 18th ACM Conference on Computer Supported Cooperative Work \& Social Computing}} \emph{(\bibinfo{series}{CSCW '15})}. \bibinfo{publisher}{Association for Computing Machinery}, \bibinfo{address}{New York, NY, USA}, \bibinfo{pages}{649–661}.
\newblock
\showISBNx{9781450329224}
\urldef\tempurl%
\url{https://doi.org/10.1145/2675133.2675277}
\showDOI{\tempurl}


\bibitem[Taylor et~al\mbox{.}(2011)]%
        {Taylor}
\bibfield{author}{\bibinfo{person}{Andrea Taylor}, \bibinfo{person}{Angus Aitken}, \bibinfo{person}{David Godden}, {and} \bibinfo{person}{Judith Colligan}.} \bibinfo{year}{2011}\natexlab{}.
\newblock \showarticletitle{Group Pulmonary Rehabilitation Delivered to the Home via the Internet: Feasibility and Patient Perception}. In \bibinfo{booktitle}{\emph{Proceedings of the SIGCHI Conference on Human Factors in Computing Systems}} (Vancouver, BC, Canada) \emph{(\bibinfo{series}{CHI '11})}. \bibinfo{publisher}{Association for Computing Machinery}, \bibinfo{address}{New York, NY, USA}, \bibinfo{pages}{3083–3092}.
\newblock
\showISBNx{9781450302289}
\urldef\tempurl%
\url{https://doi.org/10.1145/1978942.1979398}
\showDOI{\tempurl}


\bibitem[Taylor et~al\mbox{.}(1984)]%
        {Taylor84}
\bibfield{author}{\bibinfo{person}{C.~B. Taylor}, \bibinfo{person}{T. Coffey}, \bibinfo{person}{K. Berra}, \bibinfo{person}{R. Iaffaldano}, \bibinfo{person}{K. Casey}, {and} \bibinfo{person}{W.~L. Haskell}.} \bibinfo{year}{1984}\natexlab{}.
\newblock \showarticletitle{Seven-day activity and self-report compared to a direct measure of physical activity}.
\newblock \bibinfo{journal}{\emph{American Journal of Epidemiology}} \bibinfo{volume}{120}, \bibinfo{number}{6} (\bibinfo{year}{1984}), \bibinfo{pages}{818--824}.
\newblock
\urldef\tempurl%
\url{https://doi.org/10.1093/oxfordjournals.aje.a113954}
\showDOI{\tempurl}


\bibitem[Todd et~al\mbox{.}(2022)]%
        {Todd}
\bibfield{author}{\bibinfo{person}{Noah Todd}, \bibinfo{person}{Jessica Pater}, \bibinfo{person}{Jason Li}, \bibinfo{person}{Camden Bohn}, \bibinfo{person}{Brian Henriksen}, \bibinfo{person}{Jeanne Carroll}, {and} \bibinfo{person}{Fen-Lei Chang}.} \bibinfo{year}{2022}\natexlab{}.
\newblock \showarticletitle{Associations Between Psychiatric Symptoms and Cognitive Impairment in Post-Acute Sequelae of COVID}.
\newblock \bibinfo{journal}{\emph{Proceedings of IMPRS}} \bibinfo{volume}{5}, \bibinfo{number}{1} (\bibinfo{year}{2022}).
\newblock


\bibitem[Townsend et~al\mbox{.}(2021)]%
        {Townsend}
\bibfield{author}{\bibinfo{person}{Liam Townsend}, \bibinfo{person}{Joanne Dowds}, \bibinfo{person}{Kate O’Brien}, \bibinfo{person}{Grainne Sheill}, \bibinfo{person}{Adam~H. Dyer}, \bibinfo{person}{Brendan O’Kelly}, {and} \bibinfo{person}{John~P. Hynes}.} \bibinfo{year}{2021}\natexlab{}.
\newblock \showarticletitle{Persistent poor health after COVID-19 is not associated with respiratory complications or initial disease severity}.
\newblock \bibinfo{journal}{\emph{Annals of the American Thoracic Society}} \bibinfo{volume}{18}, \bibinfo{number}{6} (\bibinfo{year}{2021}), \bibinfo{pages}{997--1003}.
\newblock
\urldef\tempurl%
\url{https://doi.org/10.1513/AnnalsATS.202009-1175OC}
\showDOI{\tempurl}


\bibitem[Troiano et~al\mbox{.}(2021)]%
        {Troiano}
\bibfield{author}{\bibinfo{person}{Giovanni~M Troiano}, \bibinfo{person}{Matthew Wood}, \bibinfo{person}{Mustafa~Feyyaz Sonbudak}, \bibinfo{person}{Riddhi~Chandan Padte}, {and} \bibinfo{person}{Casper Harteveld}.} \bibinfo{year}{2021}\natexlab{}.
\newblock \showarticletitle{“Are We Now Post-COVID?”: Exploring Post-COVID Futures Through a Gamified Story Completion Method}. In \bibinfo{booktitle}{\emph{Designing Interactive Systems Conference 2021}} (Virtual Event, USA) \emph{(\bibinfo{series}{DIS '21})}. \bibinfo{publisher}{Association for Computing Machinery}, \bibinfo{address}{New York, NY, USA}, \bibinfo{pages}{48–63}.
\newblock
\showISBNx{9781450384766}
\urldef\tempurl%
\url{https://doi.org/10.1145/3461778.3462069}
\showDOI{\tempurl}


\bibitem[Vanichkachorn et~al\mbox{.}(2021)]%
        {Vanichkachorn}
\bibfield{author}{\bibinfo{person}{Greg Vanichkachorn}, \bibinfo{person}{Richard Newcomb}, \bibinfo{person}{Clayton~T. Cowl}, \bibinfo{person}{M.~Hassan Murad}, \bibinfo{person}{Laura Breeher}, \bibinfo{person}{Sara Miller}, \bibinfo{person}{Michael Trenary}, \bibinfo{person}{Daniel Neveau}, {and} \bibinfo{person}{Steven Higgins}.} \bibinfo{year}{2021}\natexlab{}.
\newblock \showarticletitle{Post–COVID-19 syndrome (long haul syndrome): description of a multidisciplinary clinic at Mayo clinic and characteristics of the initial patient cohort}.
\newblock \bibinfo{journal}{\emph{Mayo clinic proceedings}} \bibinfo{volume}{96}, \bibinfo{number}{7} (\bibinfo{year}{2021}), \bibinfo{pages}{1782--1791}.
\newblock
\urldef\tempurl%
\url{https://doi.org/10.1016/j.mayocp.2021.04.024}
\showDOI{\tempurl}


\bibitem[Vassar and Holzmann(2013)]%
        {Vassar}
\bibfield{author}{\bibinfo{person}{Matt Vassar} {and} \bibinfo{person}{Matthew Holzmann}.} \bibinfo{year}{2013}\natexlab{}.
\newblock \showarticletitle{The retrospective chart review: important methodological considerations}.
\newblock \bibinfo{journal}{\emph{Journal of educational evaluation for health professions}} \bibinfo{volume}{10}, \bibinfo{number}{0} (\bibinfo{year}{2013}), \bibinfo{pages}{12--0}.
\newblock
\urldef\tempurl%
\url{https://doi.org/10.3352/jeehp.2013.10.12}
\showDOI{\tempurl}


\bibitem[Veinot et~al\mbox{.}(2010)]%
        {Veinot10}
\bibfield{author}{\bibinfo{person}{Tiffany~C. Veinot}, \bibinfo{person}{Kai Zheng}, \bibinfo{person}{Julie~C. Lowery}, \bibinfo{person}{Maria Souden}, {and} \bibinfo{person}{Rosalind Keith}.} \bibinfo{year}{2010}\natexlab{}.
\newblock \showarticletitle{Using Electronic Health Record Systems in Diabetes Care: Emerging Practices}. In \bibinfo{booktitle}{\emph{Proceedings of the 1st ACM International Health Informatics Symposium}} (Arlington, Virginia, USA) \emph{(\bibinfo{series}{IHI '10})}. \bibinfo{publisher}{Association for Computing Machinery}, \bibinfo{address}{New York, NY, USA}, \bibinfo{pages}{240–249}.
\newblock
\showISBNx{9781450300308}
\urldef\tempurl%
\url{https://doi.org/10.1145/1882992.1883026}
\showDOI{\tempurl}


\bibitem[Vitak et~al\mbox{.}(2016)]%
        {Vitak16}
\bibfield{author}{\bibinfo{person}{Jessica Vitak}, \bibinfo{person}{Katie Shilton}, {and} \bibinfo{person}{Zahra Ashktorab}.} \bibinfo{year}{2016}\natexlab{}.
\newblock \showarticletitle{Beyond the Belmont Principles: Ethical Challenges, Practices, and Beliefs in the Online Data Research Community}. In \bibinfo{booktitle}{\emph{Proceedings of the 19th ACM Conference on Computer-Supported Cooperative Work \& Social Computing}} (San Francisco, California, USA) \emph{(\bibinfo{series}{CSCW '16})}. \bibinfo{publisher}{Association for Computing Machinery}, \bibinfo{address}{New York, NY, USA}, \bibinfo{pages}{941–953}.
\newblock
\showISBNx{9781450335928}
\urldef\tempurl%
\url{https://doi.org/10.1145/2818048.2820078}
\showDOI{\tempurl}


\bibitem[Wang et~al\mbox{.}(2008)]%
        {Wang}
\bibfield{author}{\bibinfo{person}{Taowei~David Wang}, \bibinfo{person}{Catherine Plaisant}, \bibinfo{person}{Alexander~J. Quinn}, \bibinfo{person}{Roman Stanchak}, \bibinfo{person}{Shawn Murphy}, {and} \bibinfo{person}{Ben Shneiderman}.} \bibinfo{year}{2008}\natexlab{}.
\newblock \showarticletitle{Aligning Temporal Data by Sentinel Events: Discovering Patterns in Electronic Health Records}. In \bibinfo{booktitle}{\emph{Proceedings of the SIGCHI Conference on Human Factors in Computing Systems}} (Florence, Italy) \emph{(\bibinfo{series}{CHI '08})}. \bibinfo{publisher}{Association for Computing Machinery}, \bibinfo{address}{New York, NY, USA}, \bibinfo{pages}{457–466}.
\newblock
\showISBNx{9781605580111}
\urldef\tempurl%
\url{https://doi.org/10.1145/1357054.1357129}
\showDOI{\tempurl}


\bibitem[Weiskopf and Weng(2013)]%
        {Weiskopf}
\bibfield{author}{\bibinfo{person}{Nicole~Gray Weiskopf} {and} \bibinfo{person}{Chunhua Weng}.} \bibinfo{year}{2013}\natexlab{}.
\newblock \showarticletitle{Methods and dimensions of electronic health record data quality assessment: enabling reuse for clinical research}.
\newblock \bibinfo{journal}{\emph{Journal of the American Medical Informatics Association}} \bibinfo{volume}{20}, \bibinfo{number}{1} (\bibinfo{year}{2013}), \bibinfo{pages}{144--151}.
\newblock
\urldef\tempurl%
\url{https://doi.org/10.1136/amiajnl-2011-000681}
\showDOI{\tempurl}


\bibitem[Yong(2021)]%
        {Yong}
\bibfield{author}{\bibinfo{person}{Shin~Jie Yong}.} \bibinfo{year}{2021}\natexlab{}.
\newblock \showarticletitle{Long COVID or post-COVID-19 syndrome: putative pathophysiology, risk factors, and treatments}.
\newblock \bibinfo{journal}{\emph{Infectious Diseases}} \bibinfo{volume}{53}, \bibinfo{number}{10} (\bibinfo{year}{2021}), \bibinfo{pages}{737--754}.
\newblock
\urldef\tempurl%
\url{https://doi.org/10.1080/23744235.2021.1924397}
\showDOI{\tempurl}


\bibitem[Yoo and De~Choudhury(2019)]%
        {Yoo}
\bibfield{author}{\bibinfo{person}{Dong~Whi Yoo} {and} \bibinfo{person}{Munmun De~Choudhury}.} \bibinfo{year}{2019}\natexlab{}.
\newblock \showarticletitle{Designing Dashboard for Campus Stakeholders to Support College Student Mental Health}. In \bibinfo{booktitle}{\emph{Proceedings of the 13th EAI International Conference on Pervasive Computing Technologies for Healthcare}} (Trento, Italy) \emph{(\bibinfo{series}{PervasiveHealth'19})}. \bibinfo{publisher}{Association for Computing Machinery}, \bibinfo{address}{New York, NY, USA}, \bibinfo{pages}{61–70}.
\newblock
\showISBNx{9781450361262}
\urldef\tempurl%
\url{https://doi.org/10.1145/3329189.3329200}
\showDOI{\tempurl}


\bibitem[Zhang et~al\mbox{.}(2021)]%
        {Zhang21}
\bibfield{author}{\bibinfo{person}{Zhan Zhang}, \bibinfo{person}{Karen Joy}, \bibinfo{person}{Pradeepti Upadhyayula}, \bibinfo{person}{Mustafa Ozkaynak}, \bibinfo{person}{Richard Harris}, {and} \bibinfo{person}{Kathleen Adelgais}.} \bibinfo{year}{2021}\natexlab{}.
\newblock \showarticletitle{Data Work and Decision Making in Emergency Medical Services: A Distributed Cognition Perspective}.
\newblock \bibinfo{journal}{\emph{Proc. ACM Hum.-Comput. Interact.}} \bibinfo{volume}{5}, \bibinfo{number}{CSCW2}, Article \bibinfo{articleno}{356} (\bibinfo{date}{oct} \bibinfo{year}{2021}), \bibinfo{numpages}{32}~pages.
\newblock
\urldef\tempurl%
\url{https://doi.org/10.1145/3479500}
\showDOI{\tempurl}


\bibitem[Zhu et~al\mbox{.}(2020)]%
        {Moffa}
\bibfield{author}{\bibinfo{person}{Haining Zhu}, \bibinfo{person}{Zachary~J. Moffa}, \bibinfo{person}{Xinning Gui}, {and} \bibinfo{person}{John~M. Carroll}.} \bibinfo{year}{2020}\natexlab{}.
\newblock \showarticletitle{Prehabilitation: Care Challenges and Technological Opportunities}. In \bibinfo{booktitle}{\emph{Proceedings of the 2020 CHI Conference on Human Factors in Computing Systems}} (Honolulu, HI, USA) \emph{(\bibinfo{series}{CHI '20})}. \bibinfo{publisher}{Association for Computing Machinery}, \bibinfo{address}{New York, NY, USA}, \bibinfo{pages}{1–13}.
\newblock
\showISBNx{9781450367080}
\urldef\tempurl%
\url{https://doi.org/10.1145/3313831.3376594}
\showDOI{\tempurl}


\bibitem[Zhu et~al\mbox{.}(2022)]%
        {Zhu22}
\bibfield{author}{\bibinfo{person}{Yuanda Zhu}, \bibinfo{person}{Aishwarya Mahale}, \bibinfo{person}{Kourtney Peters}, \bibinfo{person}{Lejy Matthew}, \bibinfo{person}{Felipe Giuste}, \bibinfo{person}{Blake Anderson}, {and} \bibinfo{person}{May~D. Wang}.} \bibinfo{year}{2022}\natexlab{}.
\newblock \showarticletitle{Using Natural Language Processing on Free-Text Clinical Notes to Identify Patients with Long-Term COVID Effects}. In \bibinfo{booktitle}{\emph{Proceedings of the 13th ACM International Conference on Bioinformatics, Computational Biology and Health Informatics}} (Northbrook, Illinois) \emph{(\bibinfo{series}{BCB '22})}. \bibinfo{publisher}{Association for Computing Machinery}, \bibinfo{address}{New York, NY, USA}, Article \bibinfo{articleno}{46}, \bibinfo{numpages}{9}~pages.
\newblock
\showISBNx{9781450393867}
\urldef\tempurl%
\url{https://doi.org/10.1145/3535508.3545555}
\showDOI{\tempurl}


\bibitem[Zulman et~al\mbox{.}(2016)]%
        {Zulman}
\bibfield{author}{\bibinfo{person}{Donna Zulman}, \bibinfo{person}{Nigam~H. Shah}, {and} \bibinfo{person}{Abraham Verghese}.} \bibinfo{year}{2016}\natexlab{}.
\newblock \showarticletitle{Evolutionary pressures on the electronic health record: caring for complexity}.
\newblock \bibinfo{journal}{\emph{Jama}} \bibinfo{volume}{316}, \bibinfo{number}{9} (\bibinfo{year}{2016}), \bibinfo{pages}{923--924}.
\newblock
\urldef\tempurl%
\url{https://doi.org/10.1001/jama.2016.9538}
\showDOI{\tempurl}


\end{thebibliography}

\appendix
\section{Appendix-Survey Questions}
Surveys were sent to patients every Monday morning for 12 weeks. The surveys were dynamic -- based on the previous week's answers. 

\begin{enumerate}
  \item The following are the Top 5 Post-COVID symptoms you reported to the PPCC prior to your first appointment (listed from most severe to least severe): [Insert reported-symptoms from EHR]
  \item Are these still your Top 5 symptoms? If \textbf{No}, please list your Top 5 symptoms in order from most severe to least severe. If \textbf{Yes}, is this still the most accurate ranking of your symptoms (Y/N). If \textbf{No}, please re-order your symptoms from most to least severe. 
  \item For each symptom, did they get better, stay the same, or get worse this week?
  \item During the past week, did your overall health get better, stay the same, or get worse?
  \item During your initial PPCC visit, the following recommendations were noted. For each recommendation, note whether you haven't started, making progress, completed, not going to do anything, or if the recommendation doesn't apply to you. [Insert Care plan recommendations from EHR]
  \end{enumerate}

\section{Appendix-Interview Questions}
\begin{enumerate}
  \item Do you feel like you were asked about your symptoms too much, just enough, or not enough during the study? 
  \item Did the surveys capture the information you wanted to share about your symptoms?
  \item How often would you want to update the clinic on your most severe symptoms?
  \item Did you report any of these symptoms to another doctor or provider outside of the PPCC?
  \item Was the Fitbit easy to use? Did you have any issues with using it?
  \item Did you look at your Fitbit data during the study?
  \item Are you tracking aspects of your COVID long-haul journey that we have not asked you about?
  \item Do you plan to continue wearing the Fitbit after the study? Why?
  \item How have your long-haul symptoms impacted your personal life? Professional/work life?
  \item (Share data about their sleep) -- was this similar to your pre-COVID sleep habits?
  \item (Share activity data) -- was this similar to pre-COVID activity habits?
  \item (For each survey response where care plan was "not planning on" or "doesn't apply") -- can you talk to me about why you chose this for the care plan recommendation?
\end{enumerate}

\end{document}